# A density-based topology optimization framework for steady Navier–Stokes flow with design-dependent wall actuation

Limin Chen [a,*], Yuan Liang [b], Yongbo Deng [c], Chong Wang [d,e], Youming Zhang [a,*]

[a] *School of Mechanical and Intelligent Manufacturing, Jiujiang University, Jiujiang 332005, China*
[b] *State Key Laboratory of Structural Analysis, Optimization and CAE Software for Industrial Equipment, International Research Center for Computational Mechanics, Department of Engineering Mechanics, Dalian University of Technology, Dalian 116023, China*
[c] *Institute of Microstructure Technology (IMT), Karlsruhe Institute of Technology (KIT), Karlsruhe 76344, Germany*
[d] *Changchun Institute of Optics, Fine Mechanics and Physics (CIOMP), Chinese Academy of Sciences, Changchun 130033, China*
[e] *University of Chinese Academy of Sciences, Beijing 100049, China*

**Abstract**

Design-dependent wall velocities are difficult to treat in density-based fluid topology optimization. This difficulty arises because the fluid–solid interface is represented implicitly and evolves with the design. This paper proposes a diffuse-interface formulation for topology optimization of steady incompressible Navier–Stokes flows with design-dependent wall actuation. A continuous topological description field parameterized by compactly supported radial basis functions (CS-RBFs) provides the local interface orientation, and its regularized projection defines the pseudo-density for Brinkman flow analysis. The wall velocity is imposed on a fixed mesh through an equivalent volumetric momentum source, which is localized by a diffuse-interface measure constructed from the pseudo-density and the gradient of the description field. Assuming a locally planar interface and uniformly distributed Wendland radial basis functions, the source coefficient is determined analytically by balancing the equivalent momentum input with the Brinkman resistance at the interface center. These assumptions also yield a characteristic transition-width estimate in terms of the projection steepness and support radius. Continuous adjoint analysis is performed to derive the design sensitivities. Two- and three-dimensional examples formulated using the analogy equation demonstrate the effectiveness of the proposed topology optimization method for flow-rate maximization. Body-fitted re-simulations verify the flow fields and transport performance of the optimized designs. The formulation is further extended to electroosmotic transport under the thin electric double layer (EDL) assumption using the Helmholtz–Smoluchowski relation.

* Corresponding authors.
*E-mail addresses:* 6100114@jju.edu.cn (L. Chen), 1050183228@qq.com (Y. Zhang)

## Nomenclature

***Roman symbols***

| | |
|---|---|
| $\mathbf{d}$ | prescribed actuation direction |
| $\mathbf{E}$ | electric field |
| $\mathbf{f}, \mathbf{f}_B, \mathbf{f}_{\mathrm{act}}$ | total equivalent volumetric force, Brinkman-resistance contribution, and wall-actuation contribution |
| $w_\Gamma^*$ | characteristic transition width of a locally planar diffuse interface |
| $H_\beta$ | compact regularized projection function |
| $J$ | objective functional |
| $\mathbf{n}$ | outward unit normal vector |
| $\mathbf{n}_\phi$ | regularized unit normal recovered from the description field |
| $\mathbf{P}_\phi$ | local tangential projection tensor |
| $p$ | fluid pressure |
| $\mathbf{u}^a, p^a, V^a$ | adjoint velocity, pressure, and electric potential |
| $Q$ | volumetric inlet flow rate |
| $q$ | convexity parameter in the inverse-permeability interpolation |
| $R_j$ | support radius associated with support center $j$ |
| Re | Reynolds number |
| $s_\phi$ | regularized magnitude of the description-field gradient |
| $\mathbf{u}$ | fluid velocity vector |
| $\mathbf{u}_b$ | local wall-driving velocity vector |
| $\mathbf{u}_{\mathrm{HS}}$ | Helmholtz-Smoluchowski slip velocity |
| $U_b$ | prescribed wall-actuation magnitude |
| $V$ | electric potential |

***Greek symbols***

| | |
|---|---|
| $\alpha, \alpha_{\max}$ | inverse permeability and its maximum value |
| $\kappa_T$ | analytically calibrated wall-actuation coefficient |
| $\beta$ | projection sharpness parameter |
| $\gamma$ | projected pseudo-density; unity in fluid and zero in solid |
| $\varepsilon_\phi, \varepsilon_0, \varepsilon_r$ | description-gradient regularization parameter, vacuum permittivity, and relative permittivity |
| $\zeta_0$ | zeta potential |
| $\eta$ | threshold of the compact regularized projection |
| $\theta$ | prescribed fluid-volume fraction |
| $\boldsymbol{\lambda}$ | CS-RBF coefficient vector (design variables) |
| $\lambda_D$ | Debye length |
| $\mu$ | dynamic viscosity |
| $\mu_{\mathrm{eo}}$ | electroosmotic mobility |
| $\rho$ | fluid density |
| $\sigma$ | electrical conductivity |
| $\phi$ | topological description field |
| $\delta_{\Gamma,\phi}^\beta$ | diffuse-interface measure localizing wall-induced momentum input |
| $\psi, \tilde{\boldsymbol{\Psi}}$ | compactly supported radial basis function and normalized interpolation matrix |
| $\Omega, \partial\Omega, \Gamma$ | computational domain, its external boundary, and a generic boundary segment |

***Abbreviations***

| | |
|---|---|
| CS-RBF | compactly supported radial basis function |
| EDL | electric double layer |
| GLS | Galerkin/least-squares stabilization |
| MMA | method of moving asymptotes |
| TDF | topological description field |

## 1. Introduction

Since the pioneering work of Bendsøe and Kikuchi [1], topology optimization has become an established approach for determining material layouts within prescribed design domains [2]. The principal formulations include material-distribution methods, such as solid isotropic material with penalization (SIMP) [3], bi-directional evolutionary structural optimization (BESO) [4], and discrete integer programming [5], as well as boundary-evolution methods based on level sets [6–9] and moving morphable components (MMC) [10]. Density-based formulations are particularly attractive for fluid problems because they cast channel and device design as fixed-domain material-distribution problems compatible with finite-element analysis and adjoint optimization. The porous-medium formulation of Borrvall and Petersson [11] established this paradigm for Stokes flow and was subsequently extended to channel design [12], steady incompressible Navier–Stokes flow [13], body-force-driven and unsteady flow [14–16], and large-scale computations [17]. Later developments addressed non-Newtonian flow [18,19], thermal–fluid design [20–22], porous-model accuracy and alternative discretizations [23,24], turbines and rotating machinery [25–27], surface flow [28], and turbulent flow [29–33]. Comprehensive reviews and implementation studies are available in [34,35]. Most of these formulations drive the flow through prescribed inlet conditions, pressure differences, or body forces. The optimized fluid–solid interface affects resistance and channel geometry but does not itself supply momentum to the fluid.

Wall-driven transport presents a different design problem. Surface slip [36], electroosmotic slip [37,38], electrokinetic transport [39,40], boundary-driven acoustic streaming [41], and wall-driven pumping [42] generate or modify flow through mechanisms acting at or near a solid boundary. Electroosmotic transport is particularly important in micro- and nanofluidic systems. Gregersen et al. [43] optimized dielectric distributions for induced-charge electroosmotic pumping using a diffuse-domain model that couples ionic concentration and electric-potential fields with Stokes flow. Deng et al. [44] optimized electrode patterns on prescribed microchannel walls and represented the thin-double-layer response through the Helmholtz–Smoluchowski slip relation. These studies demonstrate the value of topology optimization in electrokinetic systems, but they do not address how a prescribed wall velocity can be applied to internal fluid–solid interfaces that emerge and move during density-based topology optimization.

Alternative interface representations have been investigated in fluid topology optimization, including level-set formulations [45–51], XFEM-based approaches [52], NURBS-based isogeometric methods [53], and moving morphable void and component methods [54,55]. Their sharper geometric descriptions facilitate the evaluation of shape sensitivities and the treatment of wall-related conditions [56]. They nevertheless follow a different representation philosophy and do not eliminate the need for a fixed-domain treatment of wall actuation within the widely used density-based framework.

Design-dependent boundary effects have also been treated directly within diffuse material descriptions [57]. For structural and thermomechanical problems, boundary loads have been converted into domain integrals involving density gradients [58], while design-dependent thermal sources and thermomechanical loads have been incorporated through a NURBS-based pseudo-density representation [59]. Interface-aware constructions have also been developed for coated structures, surface layers, and narrow boundary-strip identification [60–63]. Within density-based fluid topology optimization, Bayat et al. [64] recovered implicit-wall normals from a twice-filtered design field and enforced vanishing tangential viscous stress through a volumetric penalty term in the weak Navier–Stokes formulation. The approach was later extended to turbulent wall functions [65]. These are passive wall treatments: a prescribed inlet supplies the momentum, and the recovered interface only defines a stress condition or near-wall closure. They neither introduce active momentum at the interface nor couple such input to topology evolution. Interface identification alone also does not provide the directional regularity and quantitative force calibration required for wall actuation.

Density filters, projection methods, morphology-based operators, and Helmholtz filters provide established tools for regularizing material fields and controlling length scale [66–71]. These techniques, together with the interface constructions discussed above, show that diffuse fields can provide useful boundary information. The unresolved issue is therefore not interface identification itself, but the quantitative conversion of a velocity prescribed on an evolving wall into a localized momentum input in the Brinkman equations. Such a formulation must identify the actuation region and direction, derive the forcing magnitude theoretically from the diffuse-interface and Brinkman parameters, and remain differentiable with respect to the design.

Compactly supported radial basis functions (CS-RBFs) offer the smoothness and local control needed for this purpose [72,73]. They have been used to parameterize level-set functions [74] and density fields with fewer design variables [75]. In the present setting, a continuous CS-RBF field supplies differentiable interface orientation, while its regularized projection defines the pseudo-density for Brinkman flow analysis. The support radius also provides direct control over the geometric complexity that can be represented.

Building on this CS-RBF representation, the present work converts a velocity prescribed on the evolving wall into a diffuse momentum source within the Brinkman domain. The projected field identifies the actuation layer, while the gradient of the description field provides the local geometric information required to determine the actuation direction. The source strength is derived from a locally planar-interface balance between the equivalent momentum input and Brinkman resistance. The same analysis relates the characteristic transition width to the projection steepness and RBF support radius. Combined with a continuous-adjoint formulation, this construction provides a fixed-mesh optimization framework in which the wall actuation evolves consistently with the topology.

The main contributions are summarized as follows:

1. A fixed-domain density-based formulation is developed for purely wall-driven flow topology optimization. It converts a velocity prescribed on an evolving implicit wall into a localized equivalent momentum source without explicit interface tracking or remeshing.

2. A diffuse-interface measure is constructed from the pseudo-density and the CS-RBF gradient. A local analytical balance determines the source coefficient a priori, and the associated analysis provides an estimate of the characteristic transition width.

3. The CS-RBF parameterization supplies differentiable wall orientation, reduces the number of design variables, and regulates geometric complexity through the support radius. Continuous adjoint analysis provides the design sensitivities for both the prescribed-velocity and electroosmotic formulations.

4. Two- and three-dimensional examples based on the analogy equation demonstrate flow-rate-maximizing topology optimization under wall actuation, while body-fitted re-simulations assess the resulting flow fields and transport performance. Electroosmotic examples establish the coupling among topology, electric field, interface orientation, and Helmholtz–Smoluchowski slip.

The remainder of the paper is organized as follows. Section 2 presents the governing equations, CS-RBF design representation, diffuse-interface measure, equivalent wall-actuation model, analytical parameter relations, and continuous adjoint formulation. Section 3 describes the numerical implementation and optimization procedure. Section 4 presents the prescribed-velocity and electroosmotic examples. Section 5 summarizes the principal findings and limitations.

## 2. Formulation of the proposed wall-driven flow optimization framework

### *2.1. Design-dependent wall-driven model*

Let $\Omega \subset \mathbb{R}^d$ $(d = 2,3)$ denote the fixed computational domain. The steady incompressible Navier–Stokes equations [76] are written as

$$\begin{aligned} -\mu\nabla\cdot\left(\nabla\mathbf{u}+\nabla\mathbf{u}^T\right)+\rho(\mathbf{u}\cdot\nabla)\mathbf{u}+\nabla p=\mathbf{f}, &\quad \text{in } \Omega, \\ -\nabla\cdot\mathbf{u}=0, &\quad \text{in } \Omega, \end{aligned} \tag{1}$$

where $\mathbf{u}$ and $p$ are the velocity and pressure, $\rho$ is the fluid density, $\mu$ is the dynamic viscosity, and $\mathbf{f}$ is the equivalent-domain force density. In a wall-driven problem, both the position and the local direction of the momentum input depend on the evolving fluid–solid interface, as shown in Fig. 1.

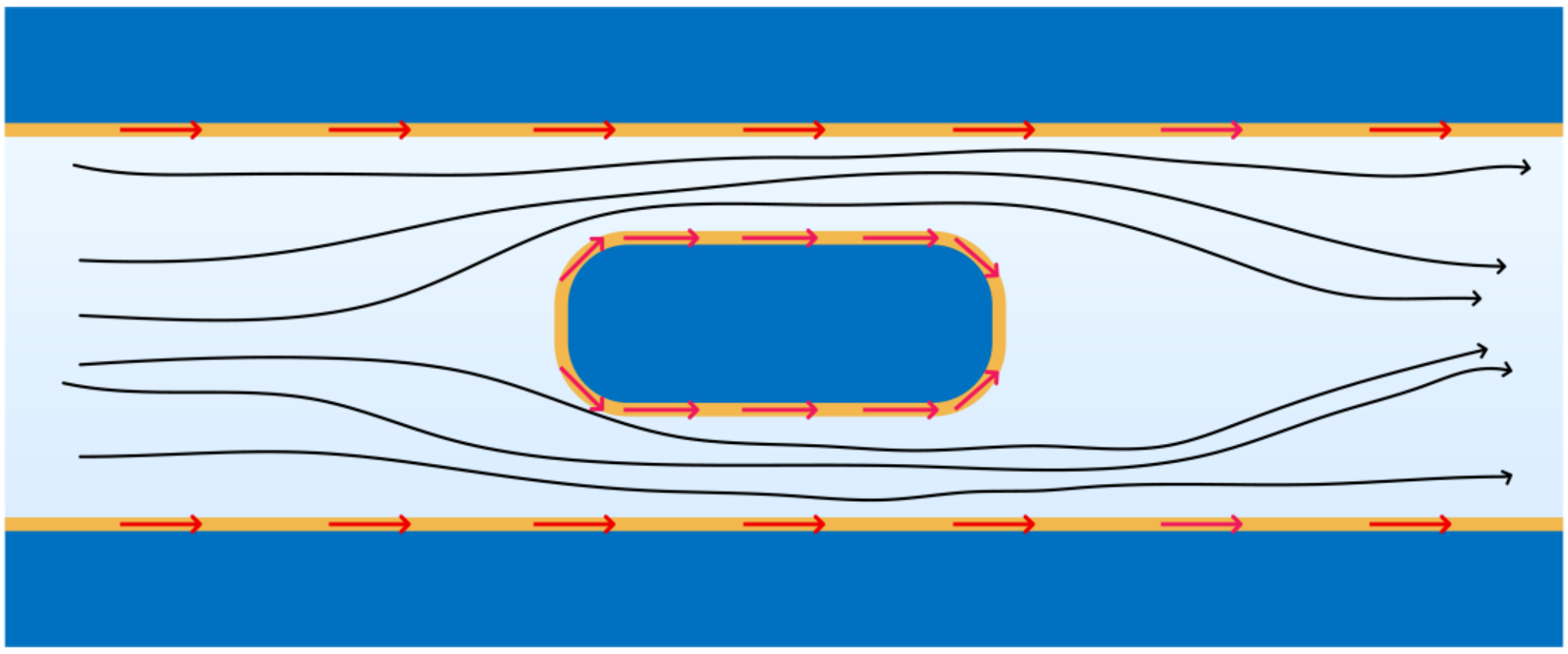

**Fig. 1.** Illustration of the flow with wall-localized forces.

Electroosmotic transport under the thin electric double layer (EDL) assumption is a representative realization. The electric body force acting on the net charge in the EDL is transmitted to the bulk liquid and represented at the outer edge of the EDL by the Helmholtz–Smoluchowski velocity, as illustrated in Fig. 2. When the wall is design dependent, the location and tangential direction of this velocity also evolve with the topology.

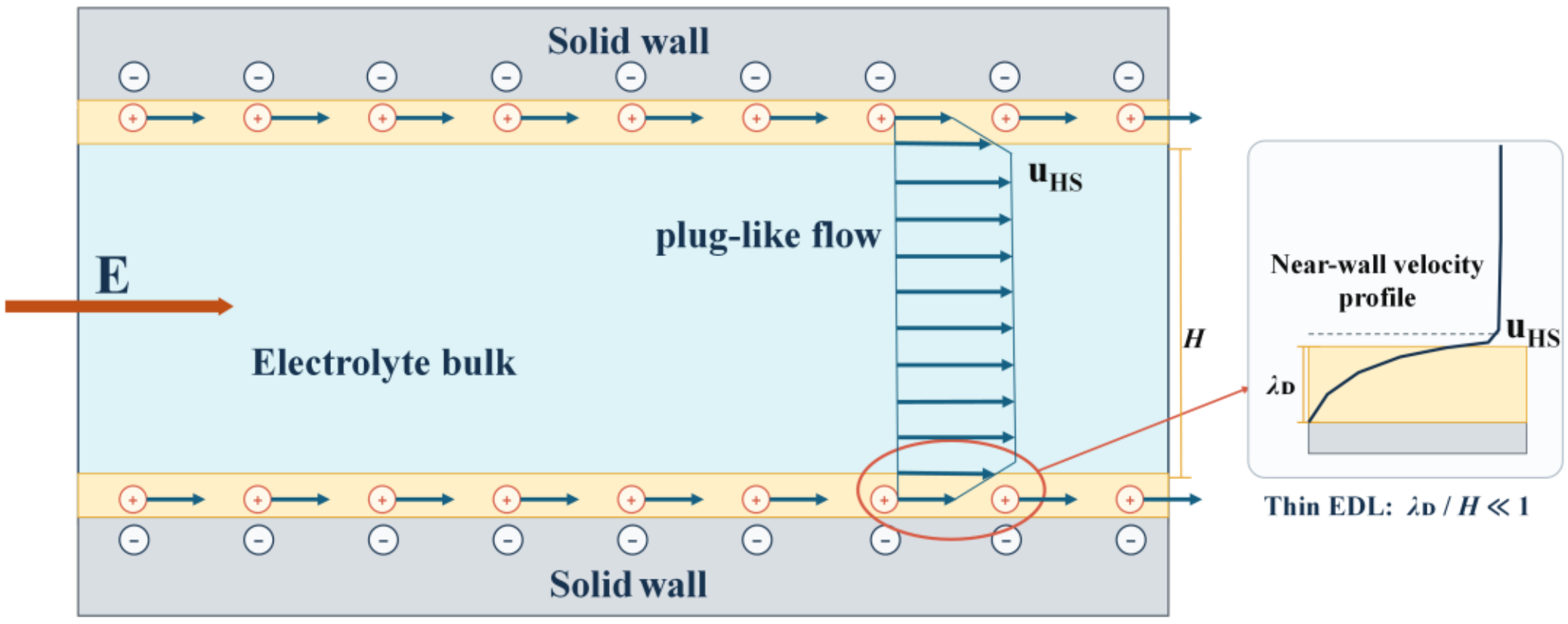

**Fig. 2.** Schematic of electroosmotic transport in a microchannel under the thin-EDL assumption. The electric force acting on the net charge within the EDL is represented at its outer edge by the Helmholtz–Smoluchowski slip velocity $\mathbf{u}_{\text{HS}}$, where $\mathbf{E}$ is the electric field, $H$ denotes the microchannel height, and $\lambda_D$ is the Debye length.

The material field is treated as a spatially varying porous medium. Following the Brinkman model, the solid phase is represented by a velocity-dependent resistance. The equivalent-domain force is written as

$$\mathbf{f} = \underbrace{-\alpha(\gamma;q)\mathbf{u}}_{\mathbf{f}_B} + \underbrace{\kappa_T \delta_{\Gamma,\phi}^{\beta} \mathbf{u}_b}_{\mathbf{f}_{\text{act}}}, \tag{2}$$

where $\kappa_T$ is the wall-actuation coefficient derived in Section 2.2.3, $\delta_{\Gamma,\phi}^{\beta}$ is the diffuse-interface measure defined in Eq. (15), and $\mathbf{u}_b$ is the prescribed local wall-driving velocity. The Brinkman term $\mathbf{f}_B$ damps the velocity toward the solid phase, whereas $\mathbf{f}_{\text{act}}$ transfers wall-induced momentum to the fluid within the transition layer.

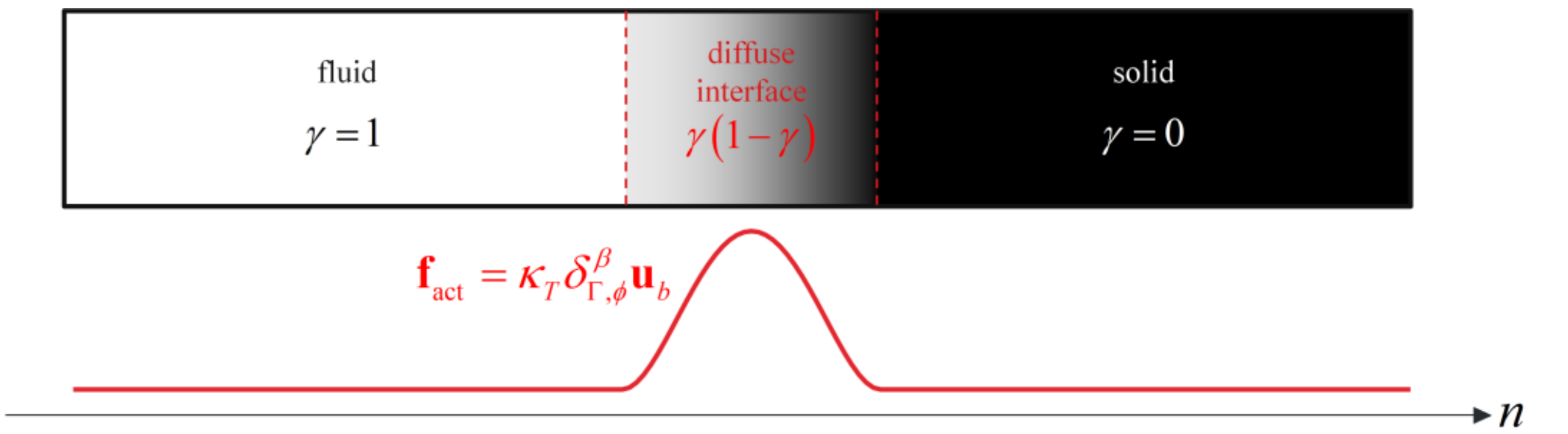

**Fig. 3.** Fixed-domain representation of wall actuation. The equivalent momentum source is localized by $\delta_{\Gamma,\phi}^{\beta}$ within the diffuse fluid–solid interface.

The external boundary is partitioned into Dirichlet and Neumann parts, $\partial\Omega = \Gamma_D \cup \Gamma_N$, with $\Gamma_D \cap \Gamma_N = \varnothing$. The boundary conditions are

$$\begin{aligned} &\mathbf{u} = \mathbf{u}_D, && \text{on } \Gamma_D, \\ &\left[-p\mathbf{I} + \mu\left(\nabla\mathbf{u} + \nabla\mathbf{u}^T\right)\right]\mathbf{n} = \mathbf{g}, && \text{on } \Gamma_N, \end{aligned} \tag{3}$$

where $\mathbf{u}_D$ and $\mathbf{g}$ are the prescribed velocity and traction, and $\mathbf{n}$ is the outward unit normal. The pseudo-density $\gamma \in [0,1]$ takes the value one in the fluid and zero in the solid. The inverse permeability [35] is interpolated as

$$\alpha(\gamma;q) = \alpha_{\max} \frac{1-\gamma}{1+q\gamma}. \tag{4}$$

The parameter $\alpha_{\max}$ controls velocity suppression in the solid phase, and $q$ controls the convexity of the interpolation. A continuation scheme decreases $q$ during optimization to promote a well-separated fluid–solid design while retaining stable early iterations. The wall-actuation coefficient derived below is evaluated from the final interpolation, for which $q=1$.

*2.2. RBF-based topology optimization framework*

*2.2.1. Topological description function parameterized by compactly supported RBFs*

A continuous topological description function (TDF) is constructed from normalized compactly supported radial basis functions (CS-RBFs) [72–75]. Let $\boldsymbol{\xi}_i\ (i=1,\ldots,M)$ denote the finite-element nodes and $\boldsymbol{\xi}_j^s\ (j=1,\ldots,N)$ the support centers. The unnormalized basis value is

$$\psi_{ij}=\psi\left(r_{ij}\right), \qquad r_{ij}=\left\|\boldsymbol{\xi}_i-\boldsymbol{\xi}_j^s\right\|_2, \tag{5}$$

where $\psi_{ij}$ is the value of the $j$th RBF evaluated at finite-element node $i$. Row-wise normalization gives

$$\tilde{\psi}_{ij}=\frac{\psi_{ij}}{\sum_{k=1}^{N}\psi_{ik}}, \tag{6}$$

and the normalized interpolation matrix is

$$\tilde{\boldsymbol{\Psi}}=\left[\tilde{\psi}_{ij}\right]_{M\times N}. \tag{7}$$

The Wendland $C^4$ function adopted in this work is

$$\psi(r)=(1-r)_+^6\left(35r^2+18r+3\right), \tag{8}$$

with the dimensionless radial coordinate

$$r=\frac{\sqrt{(x_i-x_j^s)^2+(y_i-y_j^s)^2+(z_i-z_j^s)^2}}{R_j}, \tag{9}$$

where $R_j$ is the support radius, and the $z$ term is omitted in two dimensions. The description field follows from

$$\phi=\tilde{\boldsymbol{\Psi}}\boldsymbol{\lambda}, \qquad 0\le\lambda_j\le 1. \tag{10}$$

The coefficient vector $\boldsymbol{\lambda}=[\lambda_1,\ldots,\lambda_N]^T$ contains the design variables. Because the normalized basis functions are nonnegative and form a partition of unity, $0\le\phi\le 1$. The support radius controls both the smoothness of the description field and the geometric length scale available to the optimizer.

*2.2.2. Regularized projection*

The flow analysis uses a projected pseudo-density. A compact, continuously differentiable projection [77] is adopted so that the diffuse transition occupies a prescribed interval in the description field:

$$H_\beta(\phi)=\begin{cases}0, & \phi\le\eta-\dfrac{1}{2\beta},\\ 6m^5-15m^4+10m^3, & \eta-\dfrac{1}{2\beta}<\phi<\eta+\dfrac{1}{2\beta},\\ 1, & \phi\ge\eta+\dfrac{1}{2\beta},\end{cases} \qquad m=\beta(\phi-\eta)+\frac{1}{2}. \tag{11}$$

where $\eta$ is the threshold and $\beta$ controls the projection steepness; $\eta=0.5$ is used unless otherwise stated. The pseudo-density is

$$\gamma(\phi)=1-H_\beta(\phi). \tag{12}$$

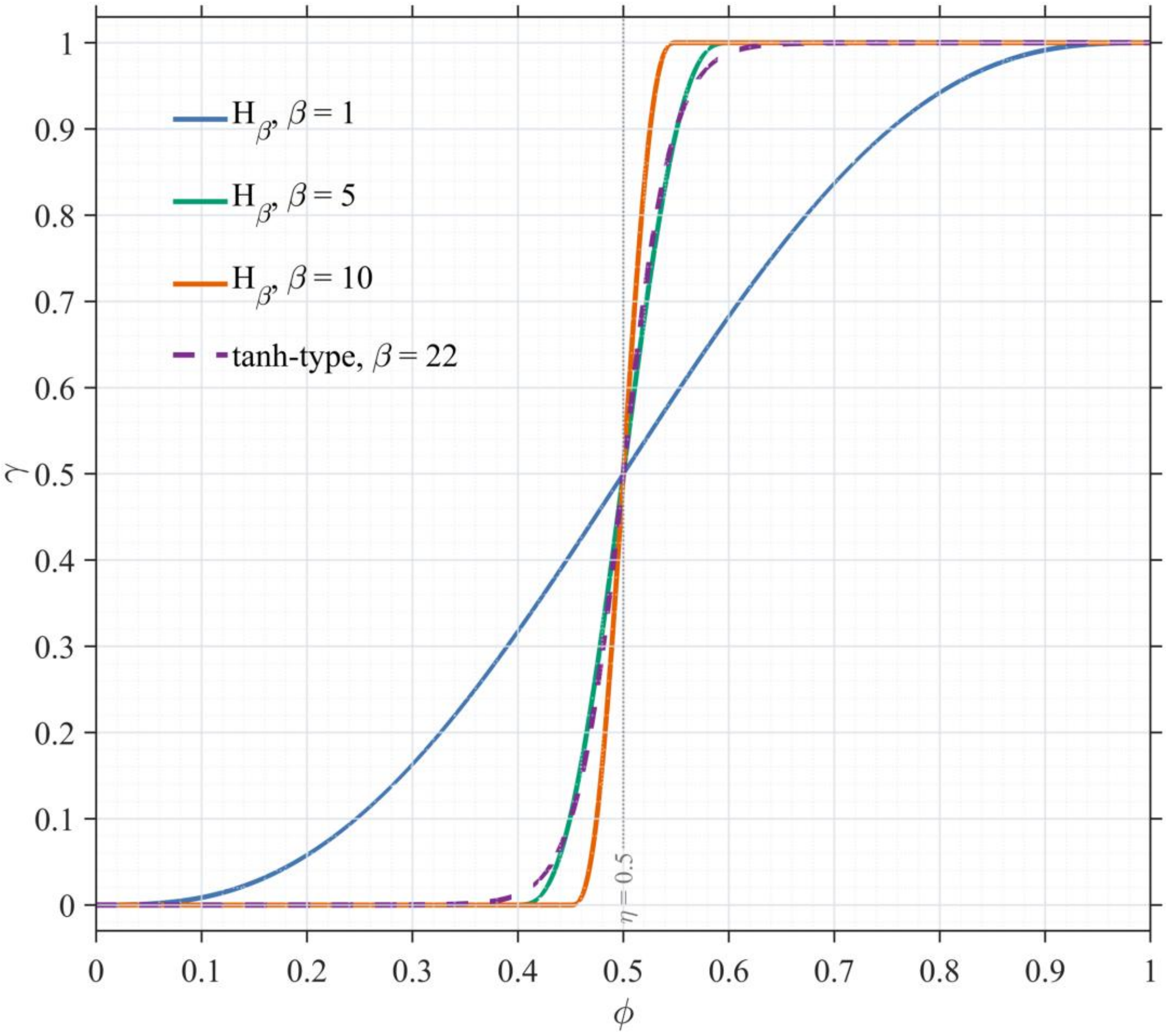

**Fig. 4.** Overlay comparison of the proposed smooth projection function and the tanh-type projection function at $\eta = 0.5$.

Its derivative, required by the sensitivity analysis, is

$$\frac{\mathrm{d}\gamma}{\mathrm{d}\phi} = \begin{cases} -30\beta m^2(1-m)^2, & |\phi - \eta| < \dfrac{1}{2\beta}, \\ 0, & \text{otherwise.} \end{cases} \tag{13}$$

For reference, the commonly used hyperbolic-tangent projection [71] is

$$H_{\tanh}(\phi) = \frac{\tanh(\beta\eta) + \tanh[\beta(\phi-\eta)]}{\tanh(\beta\eta) + \tanh[\beta(1-\eta)]}. \tag{14}$$

Unlike the tanh projection, Eq. (11) has compact support in the description-field coordinate: outside its transition interval, the projected density is exactly zero or one. This property permits a direct estimate of the physical transition width.

### *2.2.3. Diffuse-interface measure and equivalent momentum source*

The moving interface is represented by a phase weight multiplied by the gradient magnitude of the continuous description field. With a small regularization $\varepsilon_\phi > 0$, define

$$\delta^{\beta}_{\Gamma,\phi} = \gamma(1-\gamma)s_\phi, \qquad s_\phi = \sqrt{\nabla\phi \cdot \nabla\phi + \varepsilon_\phi^2}. \tag{15}$$

The factor $\gamma(1-\gamma)$ confines the measure to the projected transition, whereas $s_\phi$ converts the change in the description field into a spatial interface measure. This is the fixed-domain analogue of gradient-based representations of design-dependent surface terms [58,61].

For uniformly distributed two-dimensional Wendland $C^4$ supports with a common support radius $R_j$, the continuum approximation of a locally planar interface gives the characteristic maximum normal gradient shown in Eq. (16). The regularized projection changes over a description-field interval of $1/\beta$; hence, linearizing the description field at the interface converts this interval into a characteristic physical transition width. The complete derivation is provided in **Appendix A**.

$$s_\phi^* = \| \nabla\phi \|_\Gamma^* = \frac{16}{3\pi R_j}, \qquad w_\Gamma^* \simeq \frac{1/\beta}{s_\phi^*} = \frac{3\pi R_j}{16\beta}. \tag{16}$$

Eq. (16) shows that $R_j$ sets the smoothing scale of the CS-RBF description field, whereas $\beta$ controls the compression of that field by the projection. This expression provides a characteristic local width for an approximately planar interface represented by uniformly distributed, fully supported CS-RBFs. When these local assumptions are not satisfied, the transition width can be evaluated more directly using Eq. (A.9).

The equivalent wall-actuation coefficient is obtained from a local planar interface balance. A solid-side Brinkman boundary-layer analysis first gives the resistance associated with a prescribed interfacial velocity. The phase-weighted gradient is then normalized as a diffuse approximation of the surface Dirac measure, and the resulting source is matched to the Brinkman resistance at the center of the final projected interface. For $q = 1$, $\gamma = 1/2$, and the planar interface gradient derived in Appendix A, the intermediate normalization factors reduce to

$$\kappa_T = \frac{\alpha(1/2;1)}{\frac{1}{4} s_\phi^*} = \frac{\pi\alpha_{\max} R_j}{4}. \tag{17}$$

Thus, $\kappa_T$ is fixed by the Brinkman interpolation and the CS-RBF support radius. Substitution into the equivalent-domain force in Eq. (2) gives

$$\mathbf{f} = -\alpha(\gamma;q)\mathbf{u} + \frac{\pi\alpha_{\max} R_j}{4}\gamma(1-\gamma) s_\phi \mathbf{u}_b, \tag{18}$$

where $\mathbf{u}_b$ is the prescribed wall-driving velocity; for electroosmotic transport, $\mathbf{u}_b = \mathbf{u}_{\mathrm{HS}}$. Eq. (17) enforces the intended velocity through an ideal local balance at a planar interface center. The derivation and its assumptions are given in **Appendix B**.

For the generic examples, the prescribed velocity may follow any direction required by the actuation mechanism. When tangential actuation is considered, it is defined by

$$\mathbf{u}_b = U_b \mathbf{P}_\phi \mathbf{d}, \tag{19}$$

where $U_b$ is the prescribed magnitude and $\mathbf{d}$ is the applied driving direction. The local normal and tangent projector are

$$\mathbf{n}_\phi = \frac{\nabla\phi}{s_\phi}, \qquad \mathbf{P}_\phi = \mathbf{I} - \mathbf{n}_\phi \mathbf{n}_\phi^T. \tag{20}$$

For electroosmotic flow under the thin-EDL assumption, the Helmholtz–Smoluchowski velocity is

$$\mathbf{u}_{\mathrm{HS}} = \mu_{\mathrm{eo}} \mathbf{P}_\phi \mathbf{E}, \qquad \mu_{\mathrm{eo}} = -\frac{\varepsilon_r \varepsilon_0 \zeta_0}{\mu}, \qquad \mathbf{E} = -\nabla V, \tag{21}$$

where $\varepsilon_r$ and $\varepsilon_0$ are the relative permittivity and the permittivity of free space, respectively, $\zeta_0$ is the zeta potential, and $V$ is the electric potential. The potential satisfies

$$-\nabla \cdot \left(\sigma(\gamma)\nabla V\right) = 0 \qquad \text{in } \Omega. \tag{22}$$

The conductivity is interpolated over the fixed domain as

$$\sigma(\gamma) = \sigma_s + \left(\sigma_f - \sigma_s\right)\gamma, \tag{23}$$

where $\sigma_f$ and $\sigma_s$ are the conductivities of the electrolyte and solid phases.

### *2.3. Optimization problem*

The objective is to maximize the volumetric flow rate through the inlet boundary $\Gamma_{\mathrm{in}}$. With the outward unit normal $\mathbf{n}_{\mathrm{in}}$, the flow rate is

$$Q = -\int_{\Gamma_{\mathrm{in}}} \mathbf{u} \cdot \mathbf{n}_{\mathrm{in}} \, \mathrm{d}\Gamma. \tag{24}$$

The minimization objective is therefore

$$J = -Q = \int_{\Gamma_{\mathrm{in}}} \mathbf{u} \cdot \mathbf{n}_{\mathrm{in}} \, \mathrm{d}\Gamma. \tag{25}$$

For a prescribed wall-driving velocity, the topology optimization problem is

$$\min_{\lambda} J(\mathbf{u}, p; \boldsymbol{\lambda}), \tag{26a}$$

subject to

$$\begin{aligned} &-\mu\nabla\cdot(\nabla\mathbf{u}+\nabla\mathbf{u}^T)+\rho(\mathbf{u}\cdot\nabla)\mathbf{u}+\nabla p=-\alpha(\gamma;q)\mathbf{u}+\kappa_T\delta_{\Gamma,\phi}^{\beta}\mathbf{u}_b, \quad \text{in } \Omega, \\ &-\nabla\cdot\mathbf{u}=0, \quad \text{in } \Omega, \end{aligned} \tag{26b}$$

$$\mathbf{u}=\mathbf{u}_D, \qquad \text{on } \Gamma_D, \tag{26c}$$

$$[-p\mathbf{I}+\mu(\nabla\mathbf{u}+\nabla\mathbf{u}^T)]\mathbf{n}=\mathbf{g}, \qquad \text{on } \Gamma_N, \tag{26d}$$

$$\left|\frac{\int_\Omega \gamma \,\mathrm{d}\Omega}{\int_\Omega 1 \,\mathrm{d}\Omega}-\theta\right| \le 10^{-3}, \tag{26e}$$

$$\delta_{\Gamma,\phi}^{\beta}=\gamma(1-\gamma)s_\phi, \quad \kappa_T=\frac{\pi\alpha_{\max}R_j}{4}, \quad \gamma=1-H_\beta(\phi), \quad \phi=\tilde{\boldsymbol{\Psi}}\boldsymbol{\lambda}, \quad 0\le\boldsymbol{\lambda}\le 1, \tag{26f}$$

where $\theta\in(0,1)$ is the prescribed fluid volume fraction. For electroosmotic actuation, $\mathbf{u}_b$ is replaced by $\mathbf{u}_{\mathrm{HS}}$ and the design-dependent potential problem is added:

$$\min_{\lambda} J(\mathbf{u}, p, V; \boldsymbol{\lambda}), \tag{27a}$$

subject to

$$\begin{aligned} &-\mu\nabla\cdot(\nabla\mathbf{u}+\nabla\mathbf{u}^T)+\rho(\mathbf{u}\cdot\nabla)\mathbf{u}+\nabla p=-\alpha(\gamma;q)\mathbf{u}+\kappa_T\delta_{\Gamma,\phi}^{\beta}\mathbf{u}_{\mathrm{HS}}, \quad \text{in } \Omega, \\ &-\nabla\cdot\mathbf{u}=0, \quad \text{in } \Omega, \end{aligned} \tag{27b}$$

$$\begin{aligned} &-\nabla\cdot[\sigma(\gamma)\nabla V]=0, \quad \mathbf{E}=-\nabla V, \quad \text{in } \Omega, \\ &\mathbf{u}_{\mathrm{HS}}=\mu_{\mathrm{eo}}\mathbf{P}_\phi\mathbf{E}, \quad \text{in } \Omega, \end{aligned} \tag{27c}$$

$$\begin{aligned} &\mathbf{u}=\mathbf{u}_D \text{ on } \Gamma_D, \quad [-p\mathbf{I}+\mu(\nabla\mathbf{u}+\nabla\mathbf{u}^T)]\mathbf{n}=\mathbf{g} \text{ on } \Gamma_N, \\ &V=V_D \text{ on } \Gamma_{VD}, \quad -[\sigma(\gamma)\nabla V]\cdot\mathbf{n}=0 \text{ on } \Gamma_{VN}, \end{aligned} \tag{27d}$$

$$\left|\frac{\int_\Omega \gamma \,\mathrm{d}\Omega}{\int_\Omega 1 \,\mathrm{d}\Omega}-\theta\right| \le 10^{-3}, \tag{27e}$$

$$\delta_{\Gamma,\phi}^{\beta}=\gamma(1-\gamma)s_\phi, \quad \kappa_T=\frac{\pi\alpha_{\max}R_j}{4}, \quad \gamma=1-H_\beta(\phi), \quad \phi=\tilde{\boldsymbol{\Psi}}\boldsymbol{\lambda}, \quad 0\le\boldsymbol{\lambda}\le 1, \tag{27f}$$

The electric boundary is partitioned into prescribed-potential and insulating parts, $\Gamma_{VD}$ and $\Gamma_{VN}$, and $V_D$ denotes the prescribed electric potential.

### *2.4. Adjoint sensitivity analysis*

A continuous adjoint analysis [14,15] is used to evaluate the gradient with respect to the CS-RBF coefficients. A general steady objective functional is written as

$$J=\int_\Omega A(\mathbf{u},\nabla\mathbf{u},p,V,\nabla V;\gamma)\,\mathrm{d}\Omega+\int_{\partial\Omega} B(\mathbf{u},p,V;\gamma)\,\mathrm{d}\Gamma, \tag{28}$$

where $A$ and $B$ are the domain and boundary objective densities. The electric-potential arguments are omitted for the generic single-field examples.

#### *2.4.1. Adjoint equations*

Let $\mathbf{u}^a$ and $p^a$ be the adjoint velocity and pressure. Since the prescribed $\mathbf{u}_b$ is independent of the flow state, the flow adjoint equations are

$$-\mu\nabla\cdot[\nabla\mathbf{u}^a+(\nabla\mathbf{u}^a)^T]-\rho(\mathbf{u}\cdot\nabla)\mathbf{u}^a+\rho(\nabla\mathbf{u})^T\mathbf{u}^a+\nabla p^a=-\left[\frac{\partial A}{\partial\mathbf{u}}-\nabla\cdot\left(\frac{\partial A}{\partial(\nabla\mathbf{u})}\right)\right]-\alpha\mathbf{u}^a,\qquad \text{in }\Omega,$$
$$-\nabla\cdot\mathbf{u}^a=-\frac{\partial A}{\partial p},\qquad \text{in }\Omega, \tag{29a}$$

$$\mathbf{u}^a=-\frac{\partial B}{\partial p}\mathbf{n},\qquad \text{on }\Gamma_D, \tag{29b}$$

$$[-p^a\mathbf{I}+\mu(\nabla\mathbf{u}^a+(\nabla\mathbf{u}^a)^T)]\mathbf{n}=-\rho(\mathbf{u}\cdot\mathbf{n})\mathbf{u}^a-\frac{\partial A}{\partial(\nabla\mathbf{u})}\mathbf{n}-\frac{\partial B}{\partial\mathbf{u}},\qquad \text{on }\Gamma_N, \tag{29c}$$

For electroosmotic flow, the dependence of $\mathbf{u}_{\mathrm{HS}}$ on $V$ introduces an electric adjoint $V^a$. With $\mathbf{P}_\phi=\mathbf{I}-\mathbf{n}_\phi\mathbf{n}_\phi^T$, its equations are

$$-\nabla\cdot\left[\sigma(\gamma)\nabla V^a+\kappa_T\delta_{\Gamma,\phi}^{\beta}\mu_{\mathrm{eo}}\mathbf{P}_\phi\mathbf{u}^a\right]=-\left[\frac{\partial A}{\partial V}-\nabla\cdot\left(\frac{\partial A}{\partial(\nabla V)}\right)\right],\qquad \text{in }\Omega, \tag{30a}$$

$$V^a=0,\qquad \text{on }\Gamma_{VD},$$
$$\left[\sigma(\gamma)\nabla V^a+\kappa_T\delta_{\Gamma,\phi}^{\beta}\mu_{\mathrm{eo}}\mathbf{P}_\phi\mathbf{u}^a\right]\cdot\mathbf{n}=-\frac{\partial A}{\partial(\nabla V)}\cdot\mathbf{n}-\frac{\partial B}{\partial V},\qquad \text{on }\Gamma_{VN}. \tag{30b}$$

*2.4.2. Adjoint sensitivity*

The description field and its gradient depend linearly on the design variables:

$$\frac{\partial\phi}{\partial\boldsymbol{\lambda}}=\tilde{\boldsymbol{\Psi}},\qquad \frac{\partial(\nabla\phi)}{\partial\boldsymbol{\lambda}}=\nabla\tilde{\boldsymbol{\Psi}}. \tag{31}$$

Define the following Jacobian matrices:

$$\mathbf{J}_\gamma=-H_\beta'(\phi)\tilde{\boldsymbol{\Psi}},\qquad \mathbf{J}_s=\mathbf{n}_\phi^T\nabla\tilde{\boldsymbol{\Psi}},\qquad \mathbf{J}_n=\frac{1}{s_\phi}\mathbf{P}_\phi\nabla\tilde{\boldsymbol{\Psi}}, \tag{32a}$$

$$\mathbf{J}_\delta=(1-2\gamma)s_\phi\mathbf{J}_\gamma+\gamma(1-\gamma)\mathbf{J}_s,\qquad \mathbf{J}_\alpha=-\alpha_{\max}\frac{1+q}{(1+q\gamma)^2}\mathbf{J}_\gamma,\qquad \mathbf{J}_\sigma=(\sigma_f-\sigma_s)\mathbf{J}_\gamma, \tag{32b}$$

$$\mathbf{J}_b=-U_b[(\mathbf{n}_\phi^T\mathbf{d})\mathbf{J}_n+\mathbf{n}_\phi(\mathbf{d}^T\mathbf{J}_n)],\qquad \mathbf{J}_{\mathrm{HS}}=-\mu_{\mathrm{eo}}[(\mathbf{n}_\phi^T\mathbf{E})\mathbf{J}_n+\mathbf{n}_\phi(\mathbf{E}^T\mathbf{J}_n)], \tag{32c}$$

The derivatives $\mathbf{J}_b$ and $\mathbf{J}_{\mathrm{HS}}$ contain the explicit variation of the interface orientation. The implicit variation of the electric field is eliminated by Eq. (30). For prescribed wall actuation, the reduced objective gradient is

$$\frac{\mathrm{d}\hat{J}}{\mathrm{d}\boldsymbol{\lambda}}=\int_\Omega\frac{\partial A}{\partial\gamma}\mathbf{J}_\gamma\,\mathrm{d}\Omega+\int_{\partial\Omega}\frac{\partial B}{\partial\gamma}\mathbf{J}_\gamma\,\mathrm{d}\Gamma+\int_\Omega(\mathbf{u}^a)^T\left[\mathbf{u}\mathbf{J}_\alpha-\kappa_T\mathbf{u}_b\mathbf{J}_\delta-\kappa_T\delta_{\Gamma,\phi}^{\beta}\mathbf{J}_b\right]\mathrm{d}\Omega, \tag{33}$$

where $\hat{J}(\boldsymbol{\lambda})$ is the reduced objective obtained after substituting the state solutions into $J$. For electroosmotic actuation, the gradient becomes

$$\mathbf{G}_{\mathrm{eo}}=\int_\Omega\frac{\partial A}{\partial\gamma}\mathbf{J}_\gamma\,\mathrm{d}\Omega+\int_{\partial\Omega}\frac{\partial B}{\partial\gamma}\mathbf{J}_\gamma\,\mathrm{d}\Gamma+\int_\Omega(\mathbf{u}^a)^T[\mathbf{u}\mathbf{J}_\alpha-\kappa_T\mathbf{u}_{\mathrm{HS}}\mathbf{J}_\delta-\kappa_T\delta_{\Gamma,\phi}^{\beta}\mathbf{J}_{\mathrm{HS}}]\mathrm{d}\Omega \tag{34a}$$

$$\frac{\mathrm{d}\hat{J}}{\mathrm{d}\boldsymbol{\lambda}}=\mathbf{G}_{\mathrm{eo}}+\int_\Omega(\nabla V\cdot\nabla V^a)\mathbf{J}_\sigma\,\mathrm{d}\Omega \tag{34b}$$

*2.4.3. Flow-rate objective*

For the flow-rate objective in Eq. (25), the general densities in Eq. (28) reduce to

$$A=0,\qquad B=\begin{cases}\mathbf{u}\cdot\mathbf{n}_{\mathrm{in}}, & \text{on }\Gamma_{\mathrm{in}},\\ 0, & \text{on }\partial\Omega\setminus\Gamma_{\mathrm{in}}.\end{cases} \tag{35}$$

Consequently,

$$\frac{\partial B}{\partial \mathbf{u}} = \mathbf{n}_{\text{in}}, \quad \text{on } \Gamma_{\text{in}}, \tag{36}$$

and all other explicit derivatives of $A$ and $B$ in Eqs. (29)–(34) vanish. The volume-constraint gradient is evaluated directly as

$$\frac{\partial}{\partial \boldsymbol{\lambda}}\left(\frac{\int_\Omega \gamma \, \mathrm{d}\Omega}{\int_\Omega 1 \mathrm{d}\Omega} - \theta\right) = \frac{\int_\Omega \mathbf{J}_\gamma \, \mathrm{d}\Omega}{\int_\Omega 1 \mathrm{d}\Omega}. \tag{37}$$

## 3. Numerical implementation

The design vector is updated using the method of moving asymptotes (MMA) [78]. The procedure, summarized in Fig. 5, is as follows:

(a) Initialize the design vector $\boldsymbol{\lambda}$;

(b) Construct the description field $\phi = \tilde{\boldsymbol{\Psi}}\boldsymbol{\lambda}$ and the projected pseudo-density $\gamma = 1 - H_\beta(\phi)$;

(c) Solve the Navier–Stokes equations with Brinkman penalization and the calibrated diffuse-interface source. For electroosmotic actuation, also solve the electric-potential equation;

(d) Solve the continuous adjoint equations and evaluate the objective and constraint gradients;

(e) Update $\boldsymbol{\lambda}$ by MMA subject to $\mathbf{0} \le \boldsymbol{\lambda} \le \mathbf{1}$;

(f) Update the convexity parameter $q$ according to the continuation schedule. If the stopping criteria are not satisfied, return to step (b);

(g) Postprocess the converged design.

The continuation begins with $q = 160$ or $q = 80$. The value is reduced to $q = 40$ at iteration 50 and is subsequently decreased every 100 iterations until $q = 1$. After the final continuation value is reached, convergence is declared when

$$\frac{1}{5}\sum_{i=0}^{4} \frac{\left|J_{k-i} - J_{k-i-1}\right|}{\left|J_k\right|} \le 4 \times 10^{-5} \tag{38}$$

at iteration $k$, or when 1000 iterations have been completed. Here, $J_k$ denotes the objective value at iteration $k$.

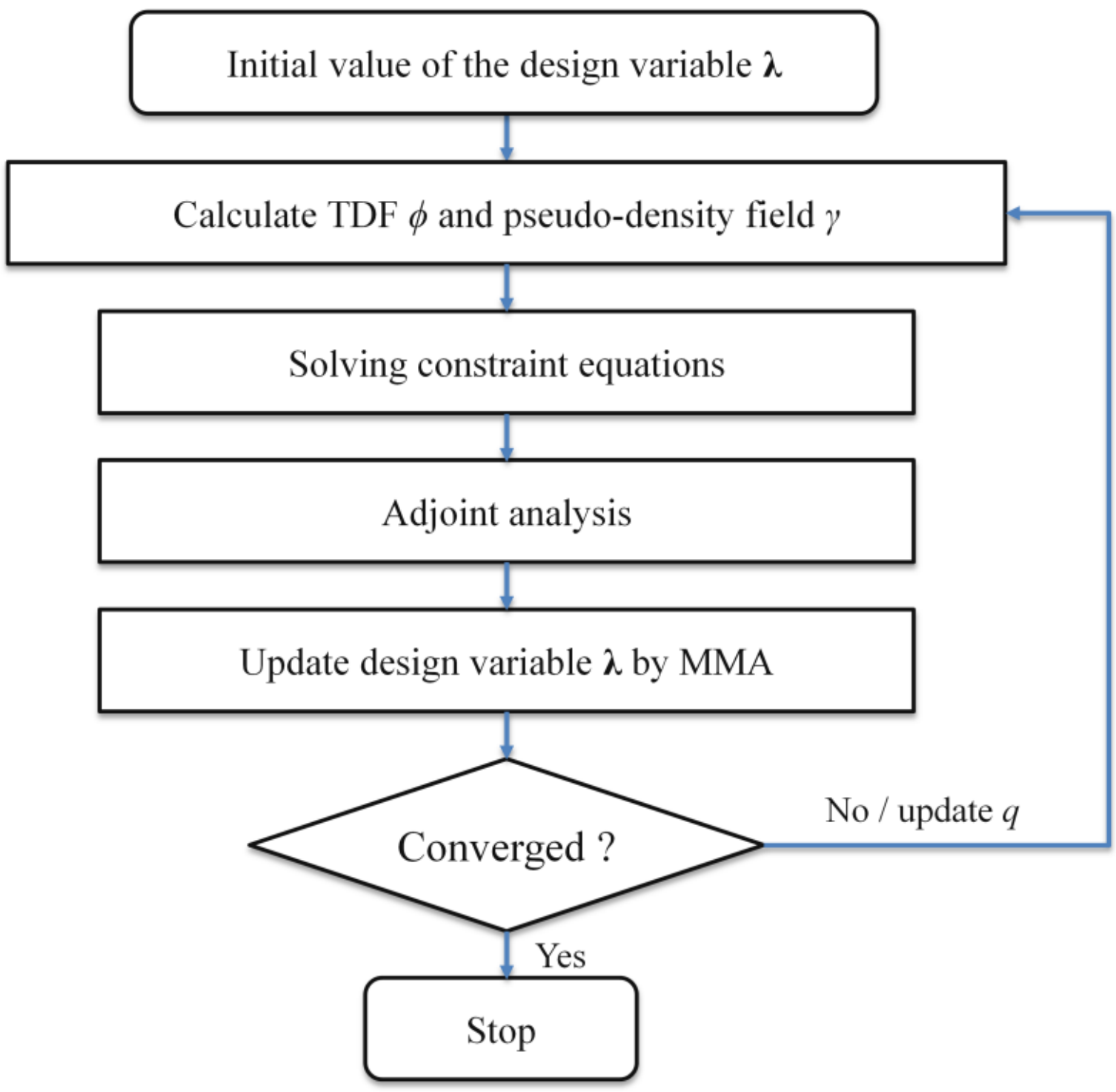

**Fig. 5.** Flowchart of the optimization procedure for the problems defined in Eqs. (26) and (27).

The nonlinear state and adjoint problems are solved at each optimization iteration using the finite-element implementation in COMSOL Multiphysics [15]. Two-dimensional structured domains use first-order quadrilateral elements, with first-order triangular elements used where an unstructured mesh is required; three-dimensional domains use first-order tetrahedral elements. Equal-order velocity–pressure interpolation is stabilized by the Galerkin/least-squares (GLS) method.

## 4. Numerical examples

This section presents a series of numerical examples to assess the proposed topology optimization framework for steady incompressible Navier–Stokes flows with design-dependent wall actuation. Unless otherwise specified, the fluid density and dynamic viscosity are both set to unity. The Reynolds number is defined as

$$\mathrm{Re} = \frac{\rho U_{\max} L}{\mu}, \tag{39}$$

where $L$ is the inlet width and $U_{\max}$ is the maximum prescribed characteristic velocity.

The numerical behavior is governed primarily by $\alpha_{\max}$, $q$, $\beta$, and the CS-RBF support radius $R_j$. The parameter $\alpha_{\max}$ controls velocity suppression in the solid phase, whereas $q$ controls the convexity of $\alpha(\gamma;q)$ and is reduced by continuation. The parameters $\beta$ and $R_j$ determine the characteristic interface thickness through Eq. (16). For each support radius, $\kappa_T$ is evaluated from Eq. (17). Unless otherwise stated, the generic settings are listed in Tables 1 and 2; the electroosmotic parameters are given separately in Section 4.6.

**Table 1**

Continuation scheme for the parameter $q$.

| Iteration number | Value of $q$ |
|---|---|
| 1–49 | 160 or 80 |
| 50–99 | 40 |
| 100–199 | 20 |
| 200–299 | 10 |
| 300–399 | 5 |
| 400–499 | 2 |
| ≥500 | 1 |

**Table 2**

Optimization parameters considered in the generic numerical examples.

| $\alpha_{\max}$ | $\beta$ | $\kappa_T$ |
|---|---|---|
| $1\times10^5$ | 5 | $\pi\alpha_{\max}R_j/4$ |

The CS-RBF support radius $R_j$, wall-actuation magnitude $U_b$, and prescribed direction $\mathbf{d}$ are specified for each benchmark. The default MMA parameters are used except for $\mathrm{asyinit}=0.05$, $\mathrm{asyincr}=1.1$, and $\mathrm{asydecr}=0.6$. Unless otherwise stated, the initial design is $\boldsymbol{\lambda}=0.5$.

The numerical studies are organized as follows. Section 4.1 compares wall-driven and inlet-velocity-driven double-pipe designs. Section 4.2 examines the effect of the CS-RBF support radius for a bending channel containing an obstacle. Section 4.3 studies the wall-actuation magnitude, driving direction, support radius, and sensitivity to controlled variations about the analytically calibrated coefficient. Sections 4.4 and 4.5 consider flux-constrained two- and three-dimensional splitters. Selected designs are reconstructed and re-simulated on body-fitted meshes. Section 4.6 applies the formulation to electroosmotic topology optimization under the thin-EDL assumption.

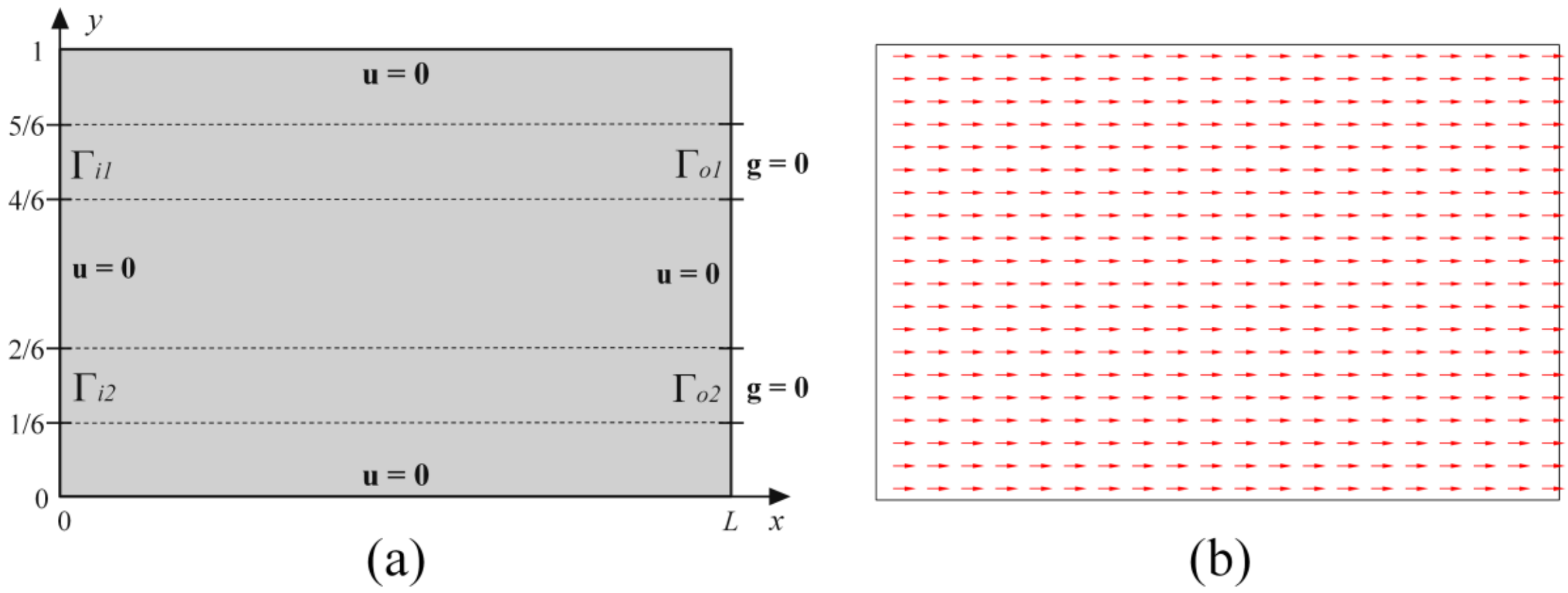

**Fig. 6.** Double-pipe benchmark for comparing wall-driven and inlet-velocity-driven topology optimization. (a) Design domain and boundary conditions; (b) prescribed driving vector $U_b\mathbf{d}$, with $U_b = 1$ and $\mathbf{d} = [1,0]^T$.

*4.1. Comparison between wall-driven and inlet-velocity-driven double-pipe flows*

The double-pipe problem compares the proposed wall-driven formulation with conventional inlet-velocity driving. In the wall-driven case, the objective is to maximize the flow rate generated by the equivalent force $\mathbf{f}$ in Eq. (2). In the inlet-driven case, $\mathbf{f} = -\alpha(\gamma;q)\mathbf{u}$, and the energy-dissipation objective is [11–13]

$$J_e = \int_\Omega \left[ \frac{\mu}{2}\nabla\mathbf{u} : \left(\nabla\mathbf{u} + \nabla\mathbf{u}^T\right) + \alpha(\gamma;q)\mathbf{u}\cdot\mathbf{u} \right] \mathrm{d}\Omega. \tag{40}$$

For a fixed pressure drop, this objective corresponds to maximizing the attainable flow rate. Therefore, the two objectives provide comparable measures of transport efficiency for the two different driving mechanisms.

The design domain is shown in Fig. 6(a). Two lengths, $L = 1.5$ and $L = 3$, are considered. For the inlet-driven formulation, the prescribed velocities are

$$\mathbf{u}_{\Gamma_{i1}} = -4\left(y - \frac{4}{6}\right)\left(\frac{5}{6} - y\right)\mathbf{n}, \qquad \mathbf{u}_{\Gamma_{i2}} = -4\left(y - \frac{1}{6}\right)\left(\frac{2}{6} - y\right)\mathbf{n}, \tag{41}$$

where $\mathbf{n}$ is the outward unit normal on $\Gamma_{i1}$ and $\Gamma_{i2}$. Open conditions are applied at the outlets. In the wall-driven formulation, the inlet and outlet are open and all remaining external boundaries are no-slip. The prescribed driving vector is $U_b\mathbf{d}$ with $U_b = 1$ and $\mathbf{d} = [1,0]^T$; projection by Eq. (19) gives the local tangential wall velocity. The initial continuation value is $q = 160$ and the fluid-volume fraction is $\theta = 1/3$. The $L = 1.5$ and $L = 3$ domains use $180\times120$ and $360\times120$ structured quadrilateral meshes, respectively. The support centers coincide with the finite-element nodes, and $R_j = 0.25$ and $R_j = 0.2$ are used for the two domains.

Fig. 7 compares the optimized results for the short benchmark with $L = 1.5$. The wall-driven formulation produces two separated, nearly straight channels, whereas the inlet-driven formulation merges the streams into a single passage. Momentum enters only through the inlets in the latter case, so reducing wall length lowers viscous loss. In the wall-driven case, the evolving interface also supplies momentum and channel separation is retained. The velocity fields in Figs. 7(c) and 7(d) show nearly flat cross-channel profiles under tangential wall actuation.

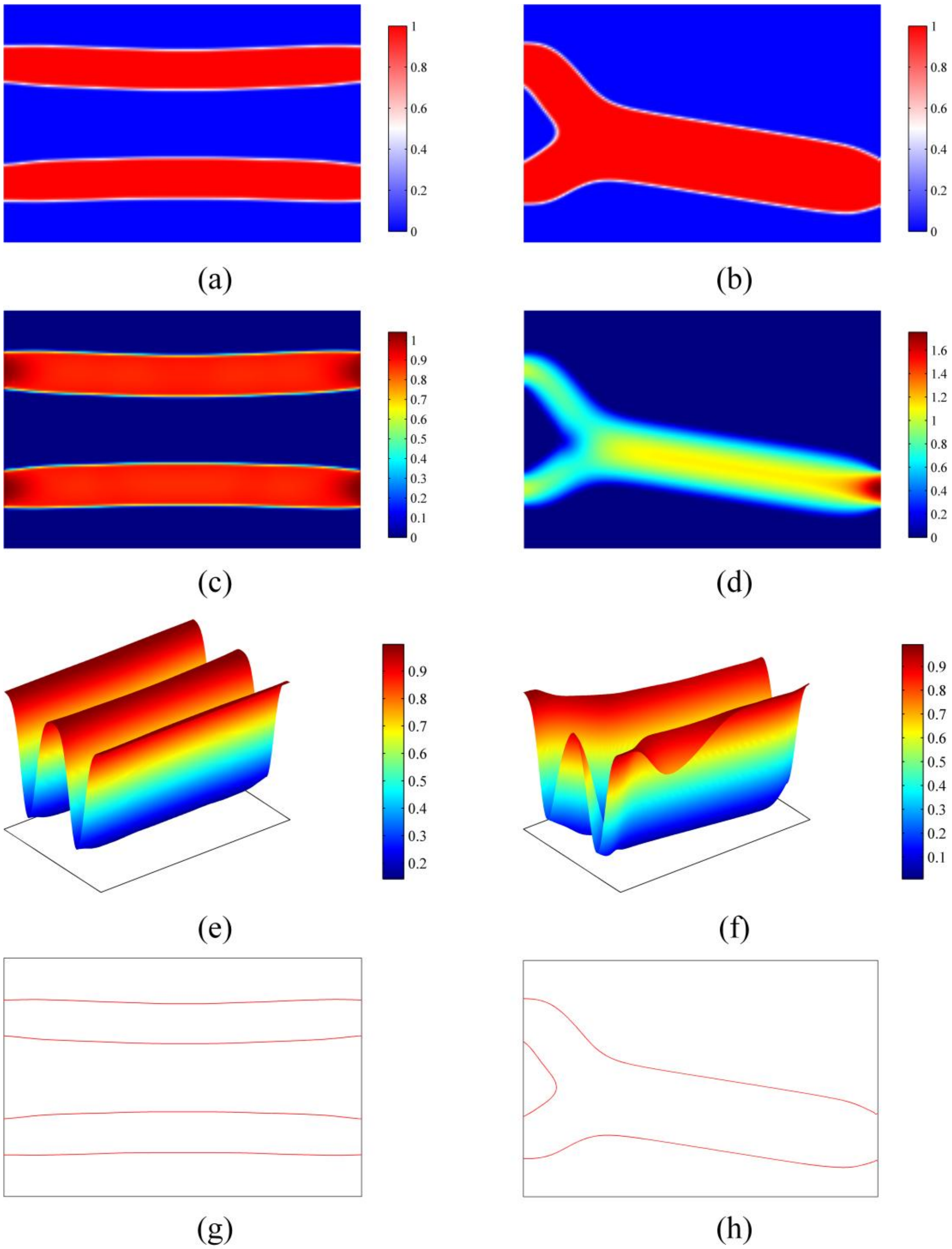

**Fig. 7.** Comparison between the wall-driven optimized results (left column) and the inlet-velocity-driven results (right column). Panels (a) and (b) show the projected pseudo-density $\gamma$; panels (c) and (d) show the velocity magnitude; panels (e) and (f) show the topological description field $\phi$; and panels (g) and (h) present the contours of $\phi = 0.5$.

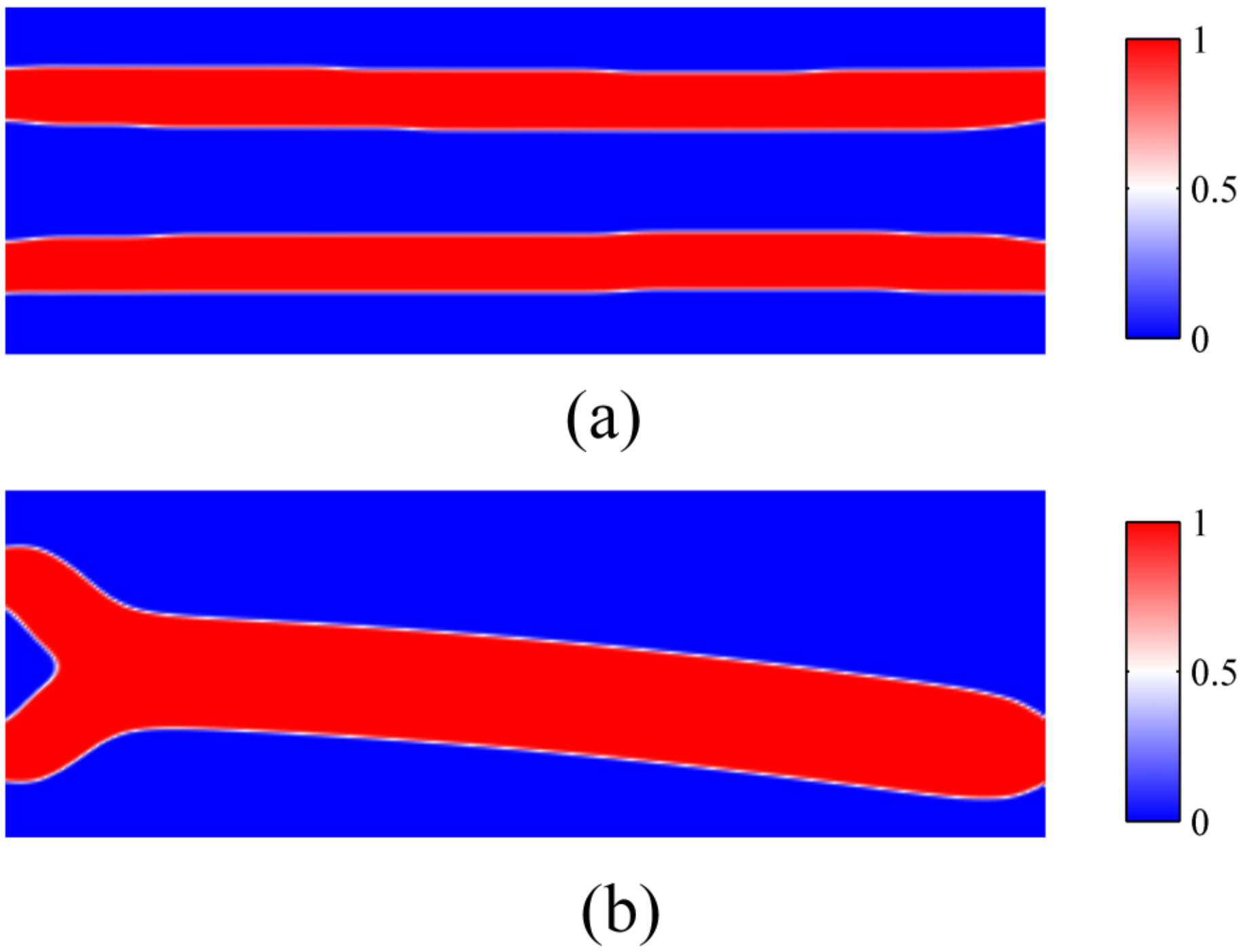

**Fig. 8.** Optimized pseudo-density fields $\gamma$ for the $L = 3$ double-pipe benchmark: (a) wall-driven; (b) inlet-velocity-driven.

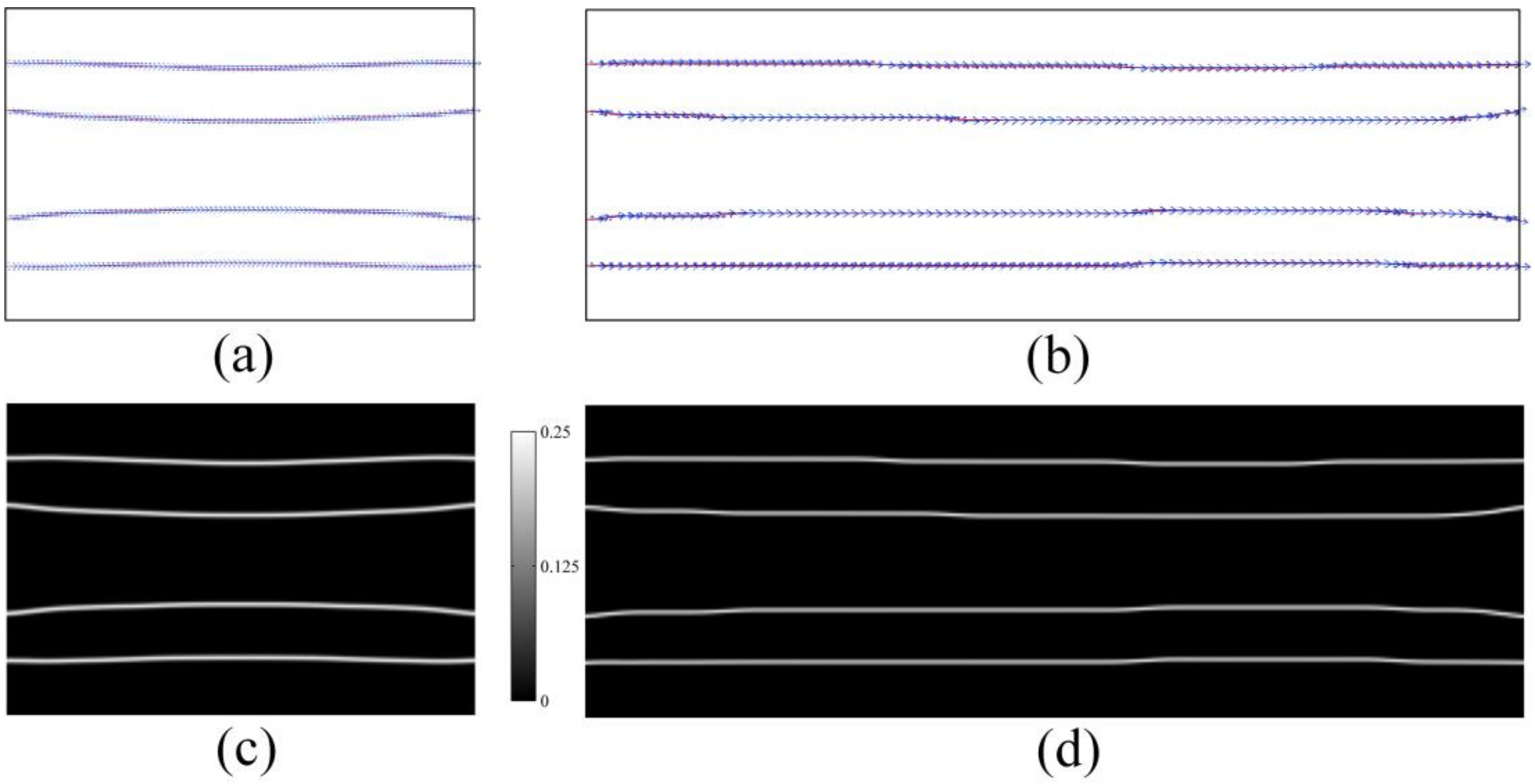

**Fig. 9.** Comparison of the optimized results for the two prescribed support radii. Panels (a) and (b) show the equivalent wall-actuation force vectors; panels (c) and (d) show the phase-weighting factor $\gamma(1-\gamma)$.

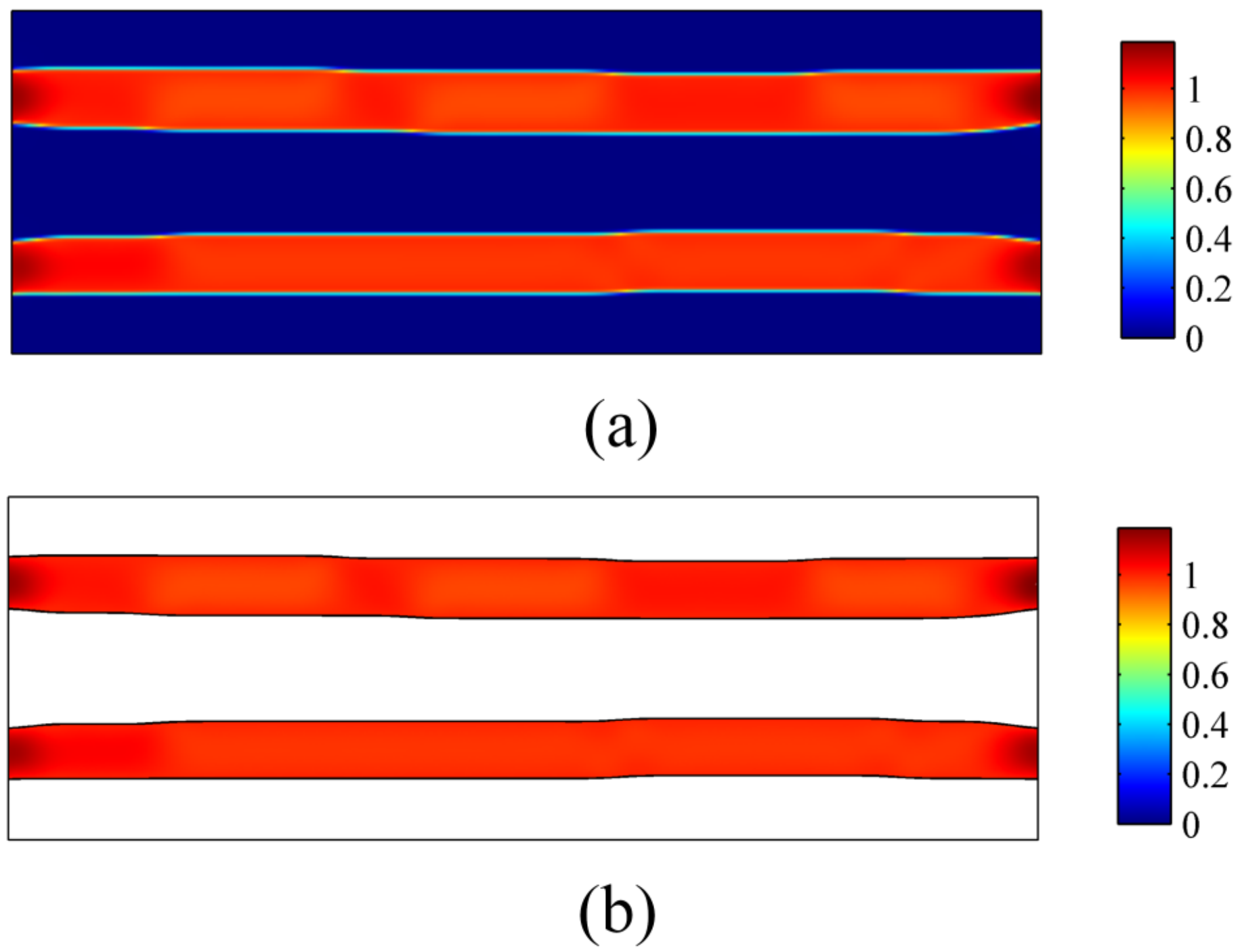

**Fig. 10.** Velocity-magnitude comparison for the $L=3$ benchmark: (a) diffuse-interface topology-optimization result; (b) body-fitted re-simulation with the corresponding tangential wall velocity.

Fig. 8 gives the optimized pseudo-density fields for $L=3$. The wall-driven design again retains two separated channels, whereas the inlet-driven design produces a merged passage. Fig. 9 shows the equivalent wall-actuation force vectors and the corresponding phase-weighting factors for the two support radii. The force remains concentrated near the evolving interfaces and is aligned with the local tangential direction. The phase weight identifies the transition region, while the complete diffuse-interface measure additionally accounts for the local description-field gradient.

A body-fitted re-simulation is performed for the wall-driven design with $L=3$. The contour $\gamma=0.5$ defines the reconstructed interface $\Gamma_w$, on which the velocity is prescribed as

$$\mathbf{u}_w = U_b\left(\mathbf{I}-\mathbf{n}_{\Gamma_w}\mathbf{n}_{\Gamma_w}^T\right)\mathbf{d} \tag{42}$$

where $\mathbf{n}_{\Gamma_w}$ is the unit normal directed from $\Gamma_w$ into the fluid. Figs. 10 and 11 compare the diffuse-interface and body-fitted solutions. The main flow paths, velocity levels, and nearly flat channel profiles agree closely. The predicted flow rates are 0.346 and 0.334, respectively, corresponding to a 3.59% difference; the remaining discrepancies are concentrated near the openings and the diffuse transition.

### *4.2. Effect of the CS-RBF support radius on bending pipe optimization with an obstacle*

This benchmark examines how the CS-RBF support radius affects the geometry and performance of a wall-driven channel optimized around an internal obstacle.

The design domain and boundary conditions are shown in Fig. 12(a). A square obstacle of side length 0.1 is fixed at the domain center as a solid non-design region. The inlet and outlet are open, whereas the external walls and obstacle are no-slip. The prescribed driving vector in Fig. 12(b) uses $U_b=1$ and $\mathbf{d}=[1,0]^T$.

A structured quadrilateral mesh with element size $h=0.01$ discretizes the domain. The obstacle is removed from the flow mesh, but CS-RBF support centers are retained on the underlying regular lattice to prevent truncation of the CS-RBF supports near the cut-out. Two radii, $R_j=0.2$ and $R_j=0.1$, corresponding to 20 and 10 element widths, are compared. The initial continuation value is $q=80$ and the fluid-volume fraction is $\theta=0.08\pi$.

The optimized results are shown in Fig. 13. The larger support radius produces a single smooth channel around the obstacle. The smaller radius permits narrow secondary branches, increasing the actuated interface and flow rate at the cost of greater geometric complexity. In both cases, $\gamma(1-\gamma)$ remains confined to the projected transition, while the full interface measure $\delta_{\Gamma,\phi}^{\beta}$ also contains the regularized gradient factor in Eq. (15).

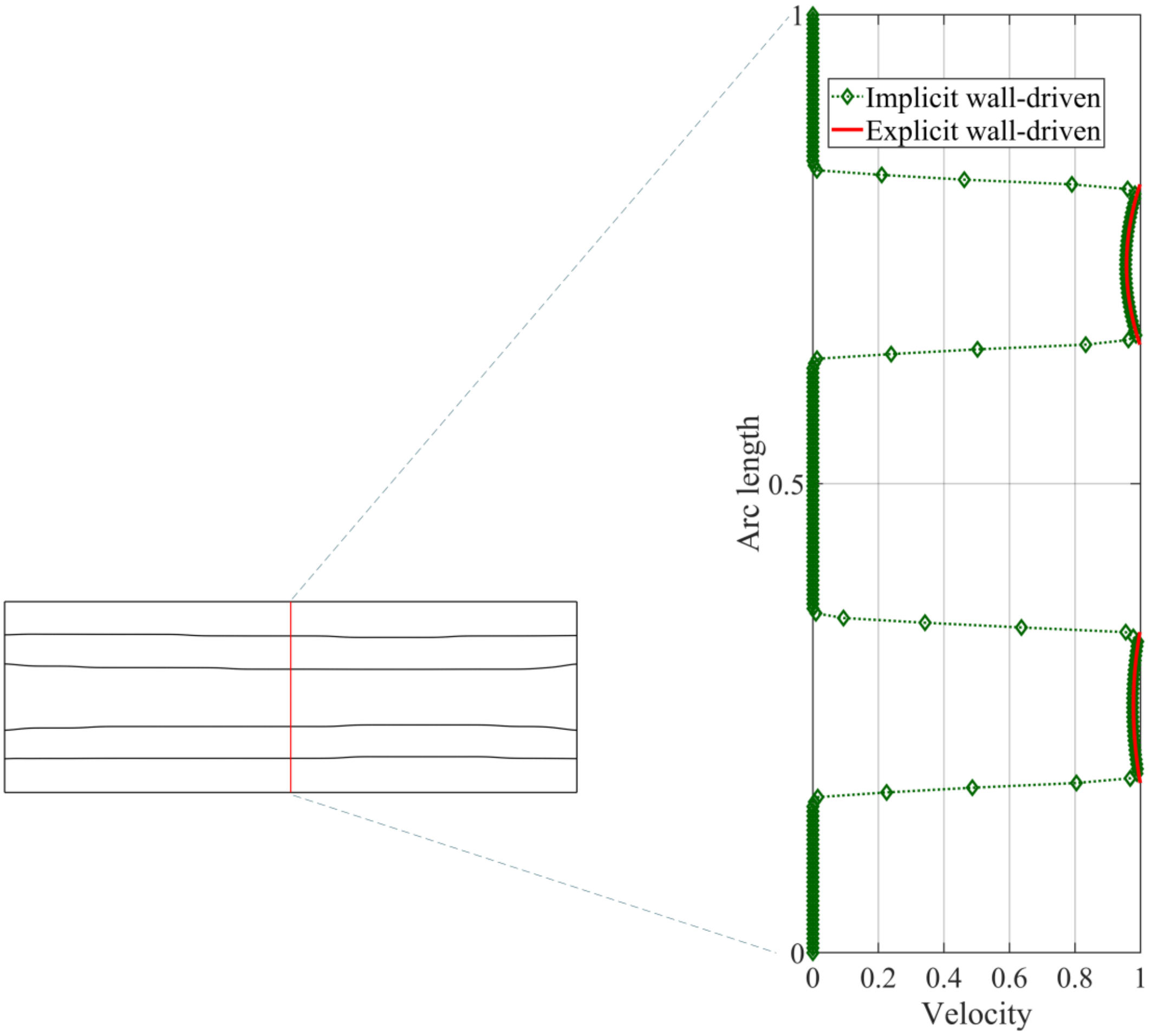

(a) (b)

**Fig. 11.** Velocity profile along a vertical sampling line for the $L = 3$ benchmark. (a) Sampling line at $x = 1.5$; (b) diffuse-interface and body-fitted velocity profiles.

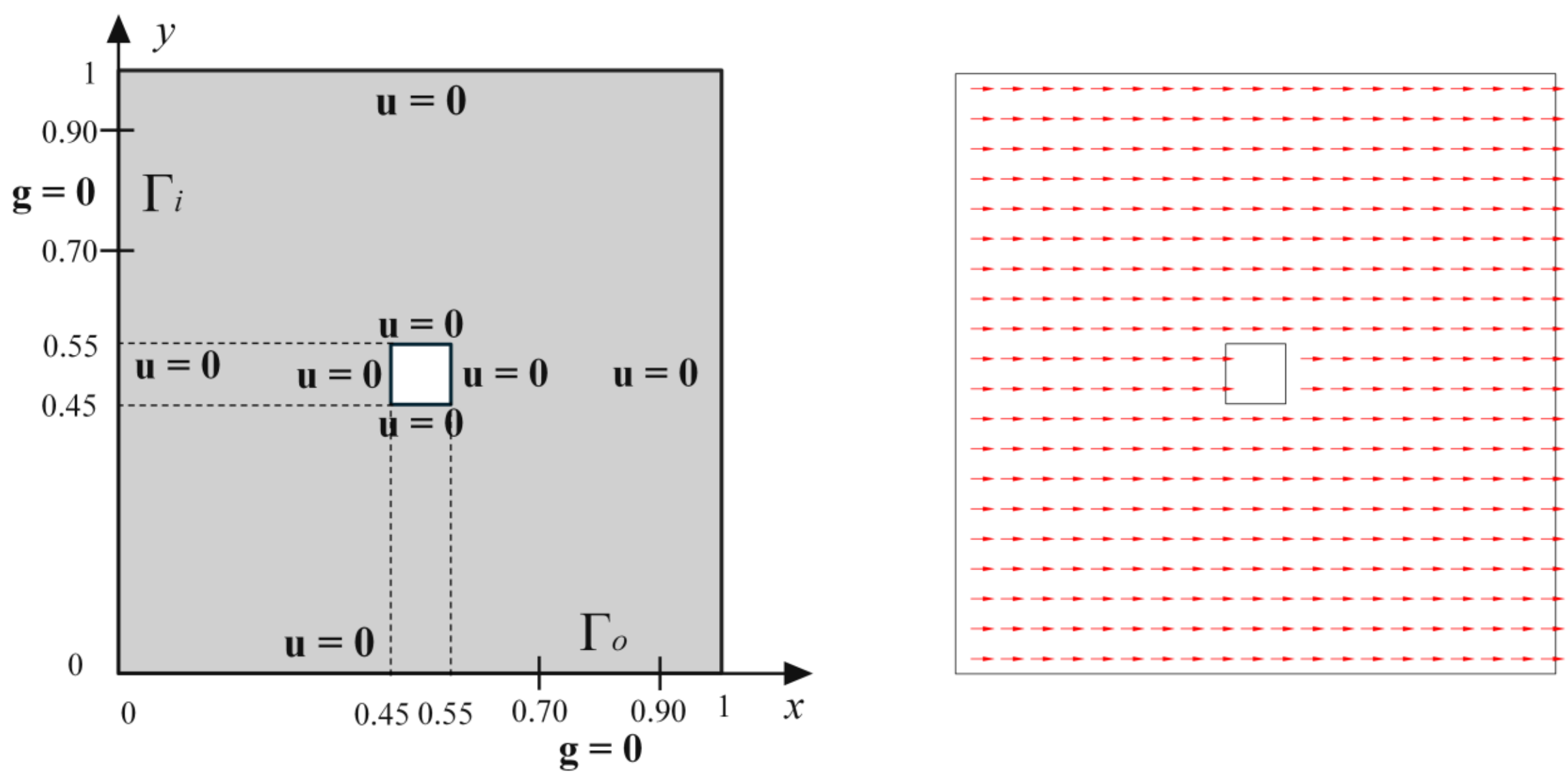

(a) (b)

**Fig. 12.** Bending-pipe benchmark with a square obstacle. (a) Design domain and boundary conditions; (b) prescribed driving vector $U_b\mathbf{d}$, with $U_b = 1$ and $\mathbf{d} = [1, 0]^T$.

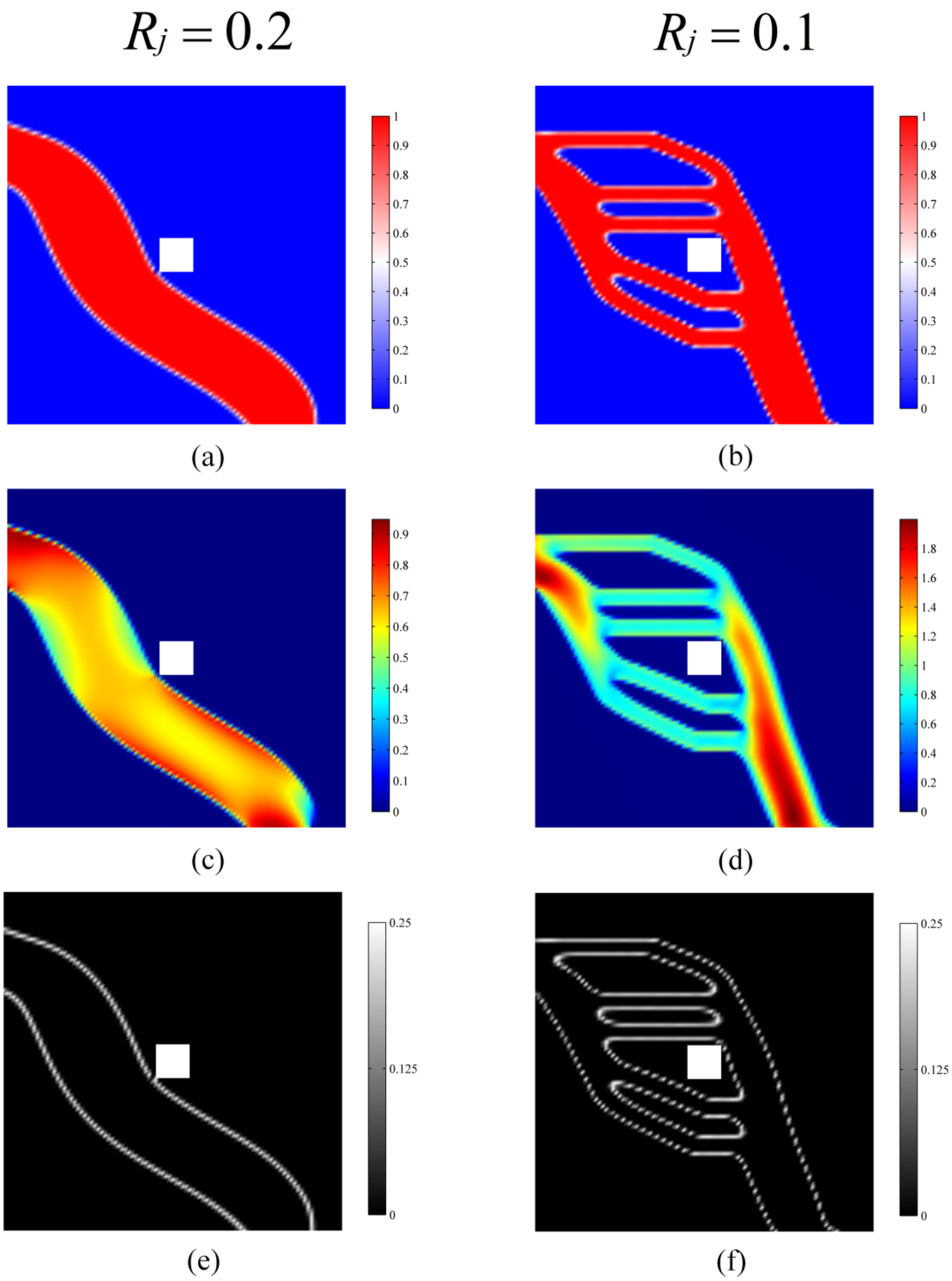

**Fig. 13.** Comparison of optimized bending-pipe designs obtained with different CS-RBF support radii (left and right columns). Panels (a) and (b) show $\gamma$; panels (c) and (d) show the velocity magnitude; and panels (e) and (f) show the phase-weighting factor $\gamma(1-\gamma)$.

### *4.3. Effects of wall-actuation magnitude, driving-field direction, CS-RBF support radius, and source-coefficient variation*

This section examines the wall-actuation magnitude $U_b$, prescribed direction $\mathbf{d}$, CS-RBF support radius $R_j$, and controlled variations of the calibrated coefficient $\kappa_T$.

The inlet and outlet in Fig. 14 are open, and the remaining external boundaries are no-slip. The initial continuation value is $q=80$ and the fluid-volume fraction is $\theta=1/3$. A $200\times150$ structured quadrilateral mesh is used, with the support centers placed at its nodes. Unless otherwise stated, $R_j=0.2$.

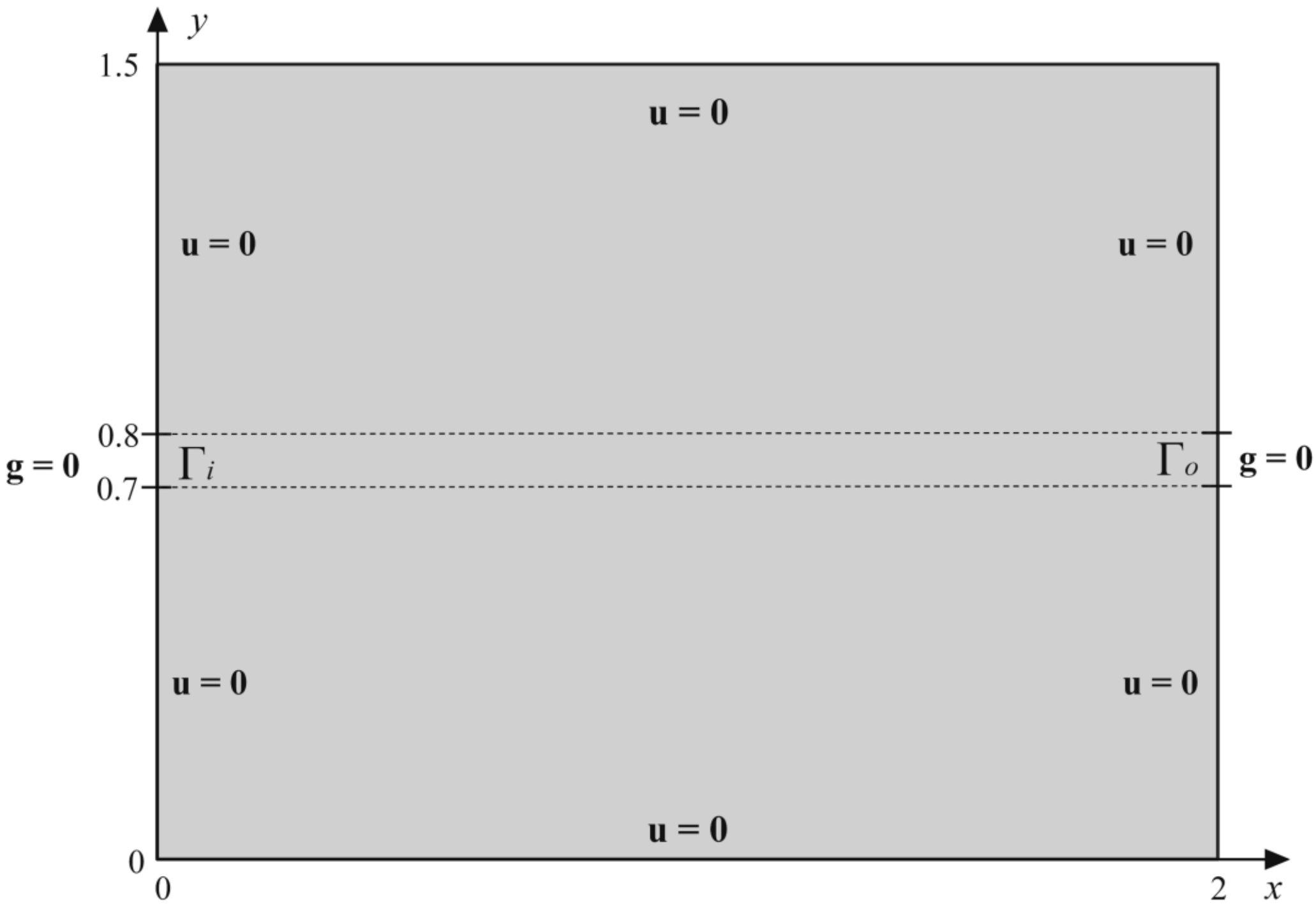

**Fig. 14.** Design domain and boundary conditions for the single-inlet–single-outlet wall-driven fluid-network optimization problem.

Fig. 15 compares three prescribed wall-actuation magnitude distributions under the same driving direction. The uniform case produces a nearly symmetric multichannel network; the piecewise case shifts the upstream collection region toward the more strongly actuated side; and the vertically varying case forms a loop-like network around a central void. In every case, the phase weight remains localized to the evolving transition region.

Fig. 16 uses the spatially varying radii $R_j = 0.2$, $0.18$, $0.12$, and $0.1$ from left to right, with $U_b = 1$ and $\mathbf{d} = [1, 0]^T$. Larger upstream radii yield wider, smoother branches, whereas smaller downstream radii permit finer features.

The prescribed direction in Fig. 17 varies spatially as

$$\mathbf{d} = \left[ \frac{1}{\left(0.5\pi\cos(\pi x)\right)^2 + 1}, \frac{0.5\pi\cos(\pi x)}{\left(0.5\pi\cos(\pi x)\right)^2 + 1} \right]^T \tag{43}$$

With $U_b = 1$, the radii $R_j = 0.1$, $0.15$, and $0.2$ give flow rates of 0.490, 0.402, and 0.328, respectively. The channels follow the prescribed field, whereas increasing $R_j$ suppresses small branches.

Fig. 18 examines $0.1\kappa_T$, $\kappa_T$, and $10\kappa_T$ under the same prescribed field. The channel paths and branch counts remain nearly unchanged, whereas the velocity magnitude scales with the coefficient. Equation (17) is used in all other examples.

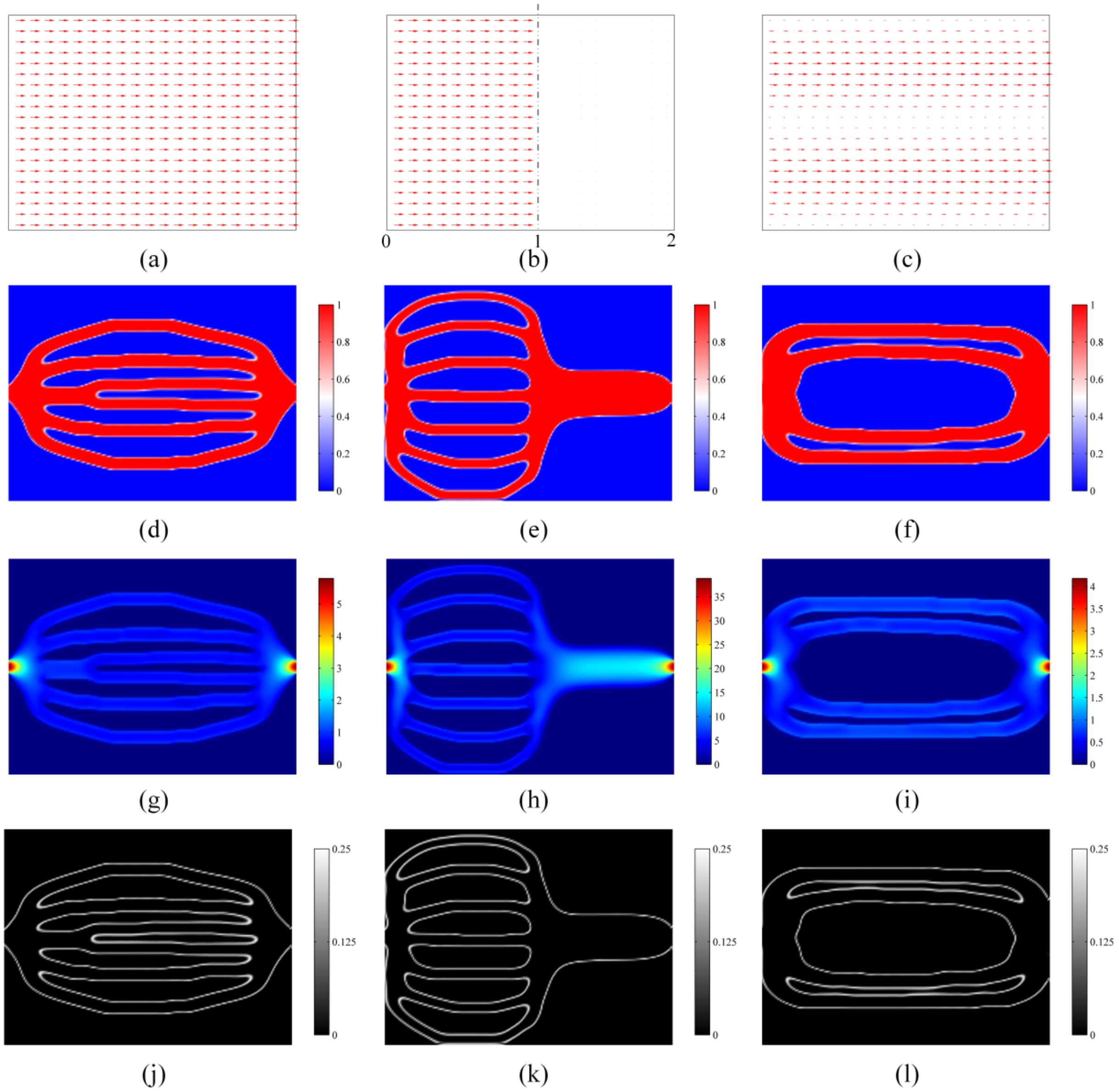

**Fig. 15.** Comparison of optimized results for three prescribed wall-actuation magnitude distributions. Panels (a)–(c) show the prescribed magnitude; panels (d)–(f) show $\gamma$; panels (g)–(i) show the velocity magnitude; and panels (j)–(l) show the phase-weighting factor $\gamma(1-\gamma)$.

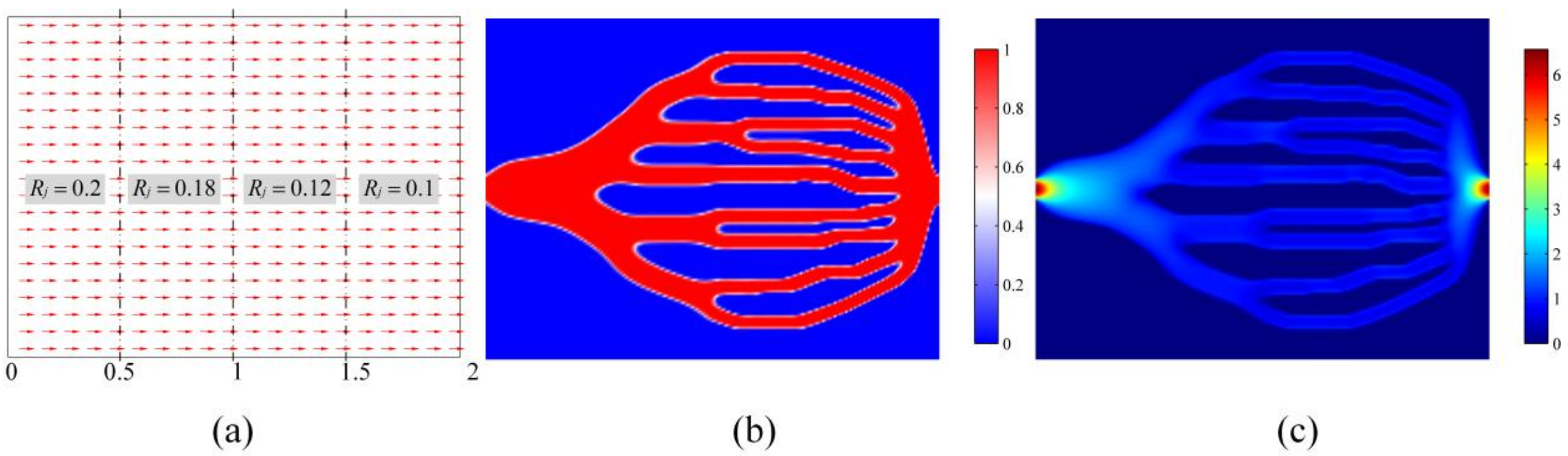

**Fig. 16.** Optimized network with spatially varying CS-RBF support radii. (a) Prescribed $U_b\mathbf{d}$ and support-radius regions, with $U_b = 1$ and $\mathbf{d} = [1, 0]^T$; (b) $\gamma$; (c) velocity magnitude.

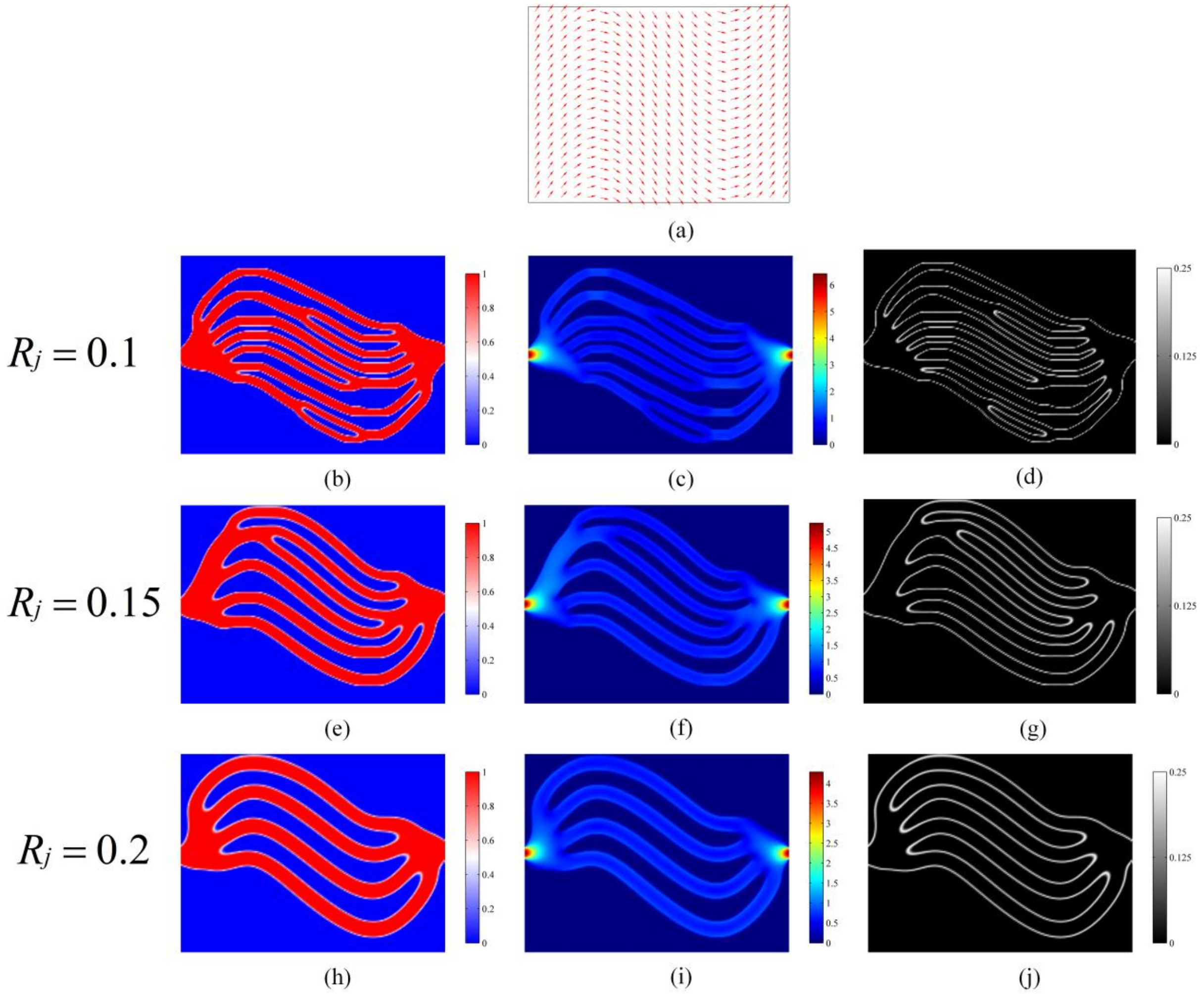

**Fig. 17.** Optimized results obtained with different CS-RBF support radii under a spatially varying driving direction. Panel (a) shows the prescribed driving field; panels (b), (e), and (h) show $\gamma$; panels (c), (f), and (i) show the velocity magnitude; and panels (d), (g), and (j) show the phase-weighting factor $\gamma(1-\gamma)$.

Fig. 19 combines $U_b = \sin(4\pi y / 3)$ with the direction in Eq. (43). The optimized network follows the resulting nonuniform actuation field.

Fig. 20 shows that the principal channel network forms by iteration 100; subsequent updates refine the boundaries and reduce intermediate densities. The equivalent wall-actuation source concentrates along the emerging interfaces, while the description field remains smooth throughout the optimization.

Fig. 21 shows rapid initial improvement followed by convergence. Each reduction in $q$ temporarily lowers the flow rate by increasing the resistance of intermediate-density regions. The phase weight evolves into narrow layers, while Eq. (16) supplies the corresponding planar-interface width estimate.

For validation, the contour $\phi = 0.5$ is reconstructed and discretized by a body-fitted triangular mesh. Fig. 22 compares the diffuse-interface result with the body-fitted solution under the corresponding velocity $\mathbf{u}_w$. The principal paths and velocity levels agree closely; residual differences occur mainly near openings and the transition layer.

Fig. 23 further shows consistent nonzero velocity intervals and peak locations along $x = 1$. Differences arise mainly where the line crosses the diffuse transition.

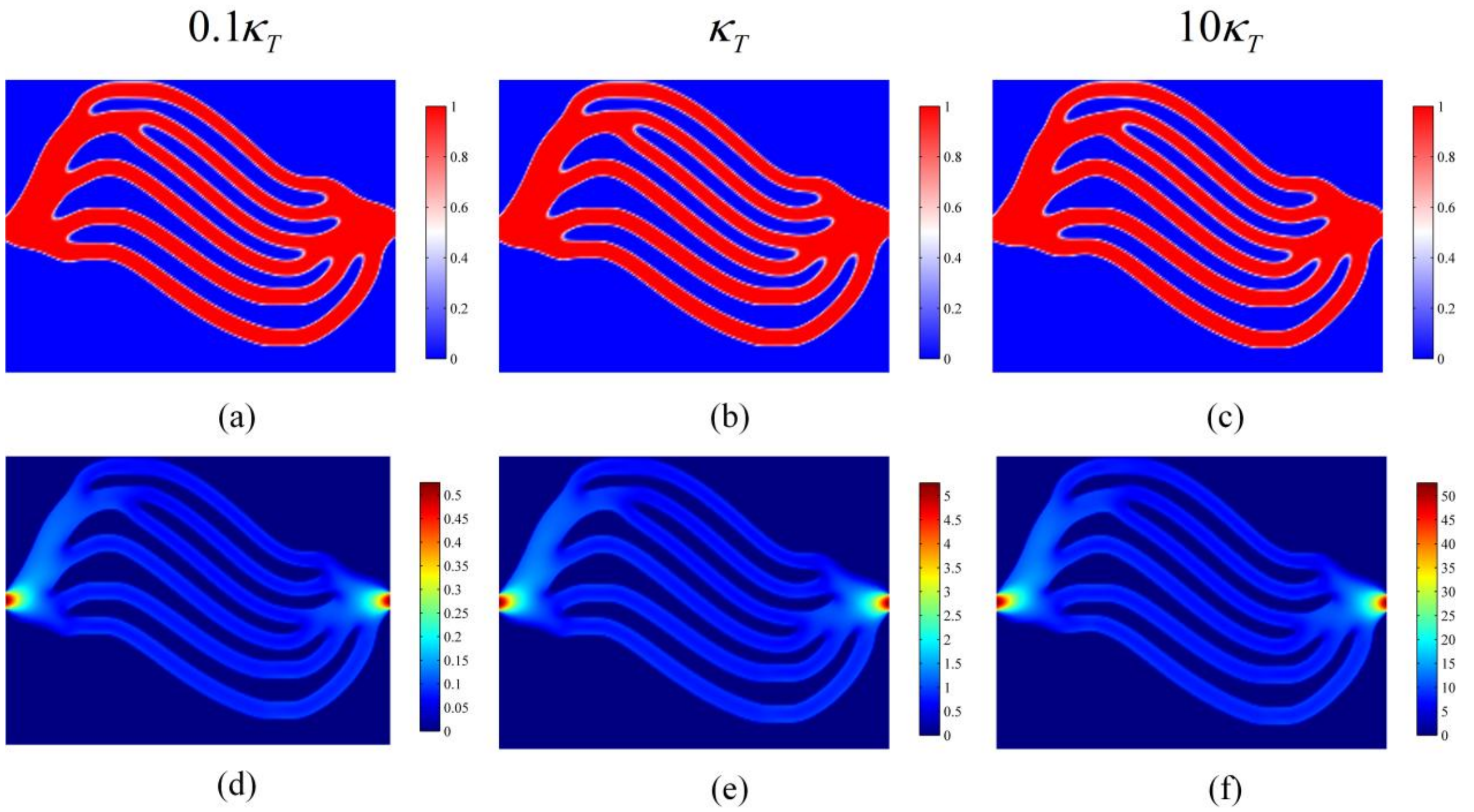

**Fig. 18.** Sensitivity of the optimized results to the source coefficient: $0.1\kappa_T$ (left column), $\kappa_T$ (middle column), and $10\kappa_T$ (right column). Panels (a)–(c) show $\gamma$, and panels (d)–(f) show the velocity magnitude.

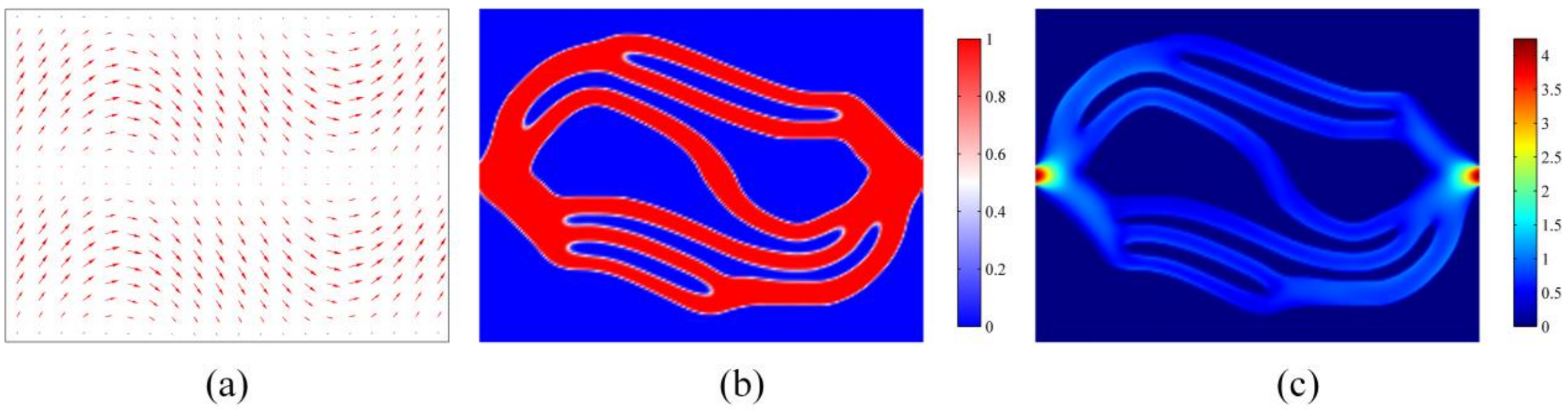

**Fig. 19.** Optimized network under spatially varying magnitude and direction. (a) Prescribed vector $U_b\mathbf{d}$ with $U_b = \sin(4\pi y/3)$ and $\mathbf{d}$ from Eq. (43); (b) $\gamma$; (c) velocity magnitude.

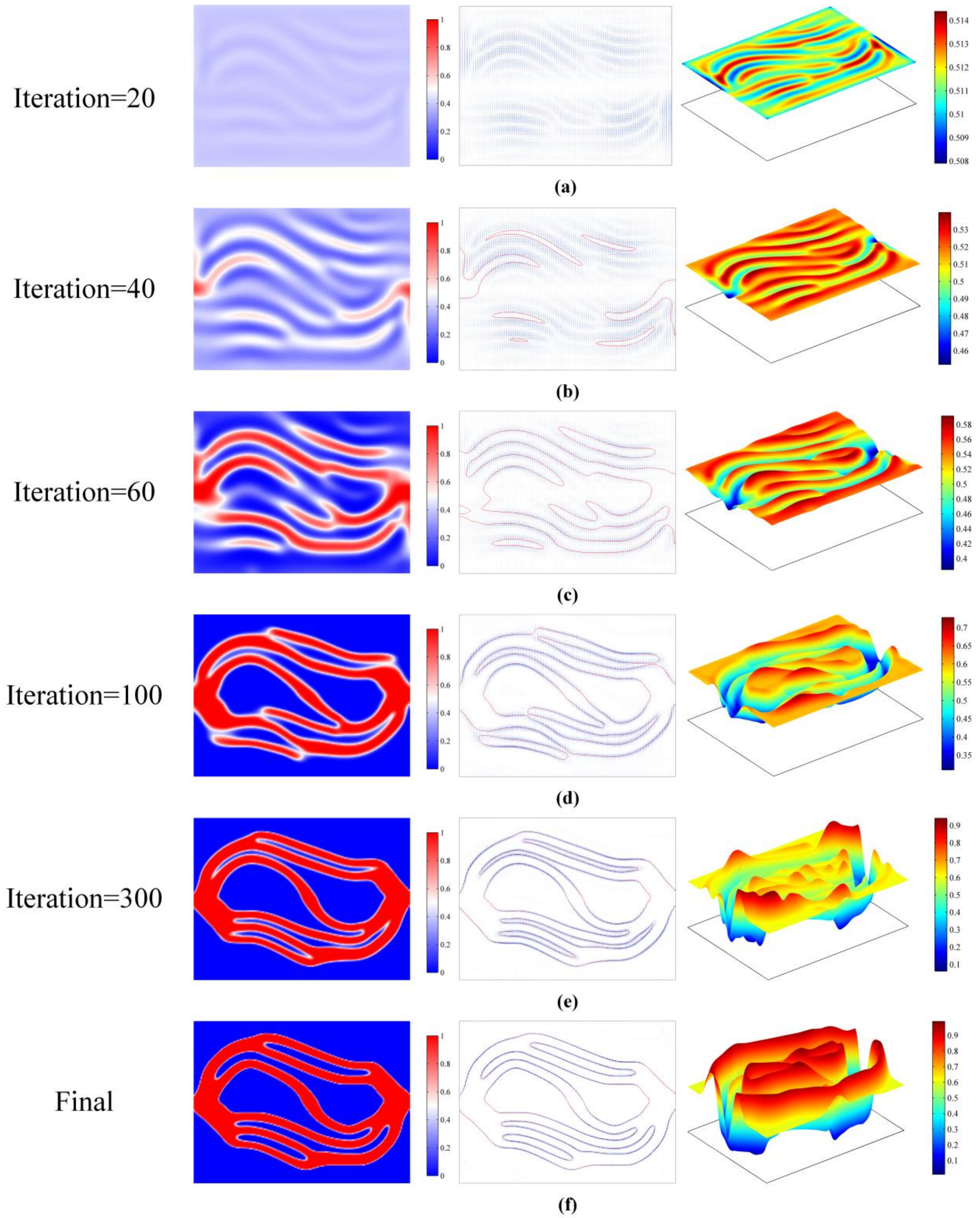

**Fig. 20.** Optimization history for Fig. 19. Rows show iterations 20, 40, 60, 100, 300, and the final design; columns show $\gamma$, the equivalent force vectors, and $\phi$.

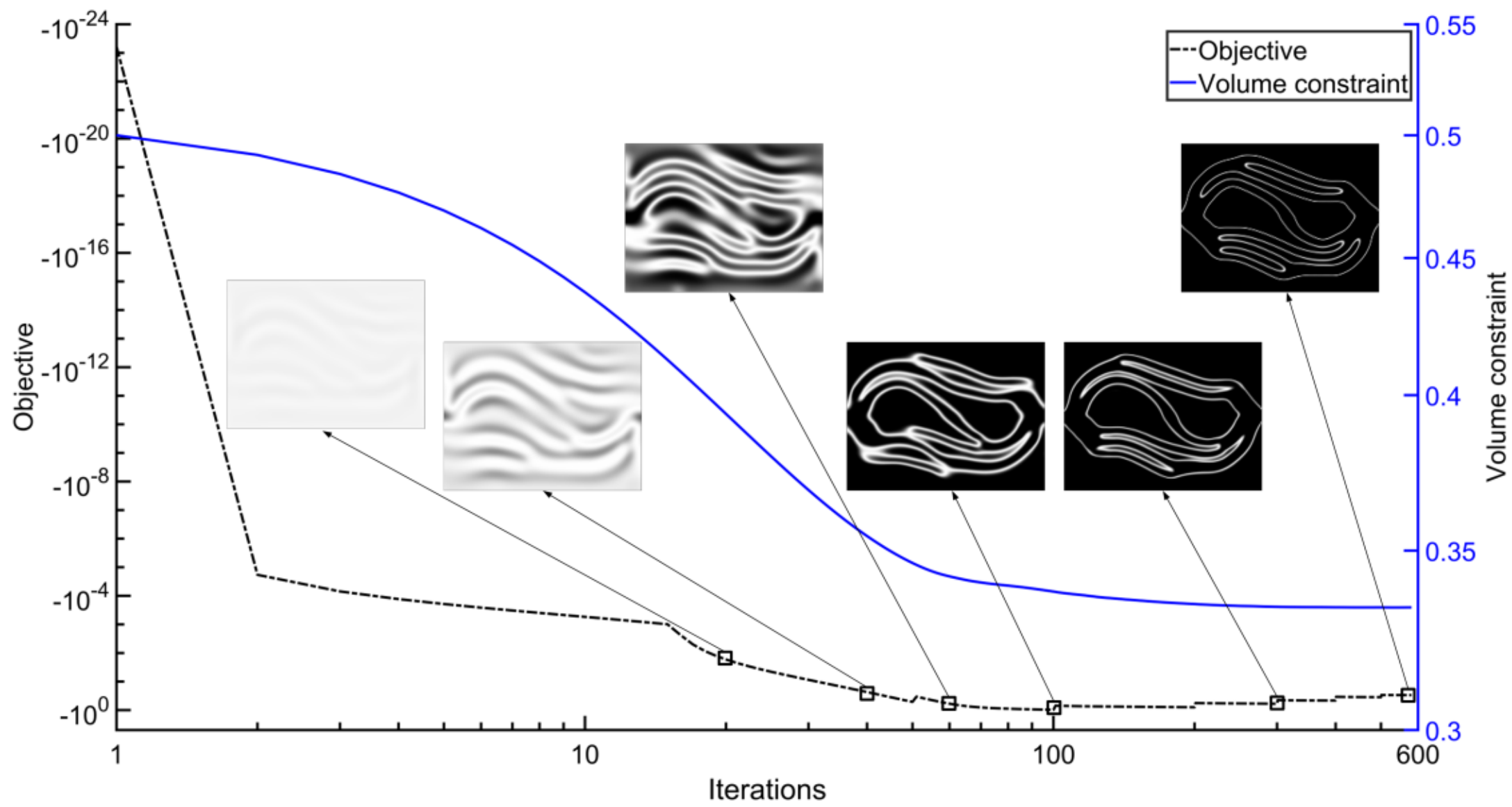

**Fig. 21.** Convergence histories of the objective value and volume constraint for the optimized topology in Fig. 19. Both the objective value and iteration number are plotted on logarithmic scales. The insets show the phase-weighting factor $\gamma(1-\gamma)$ at selected iterations.

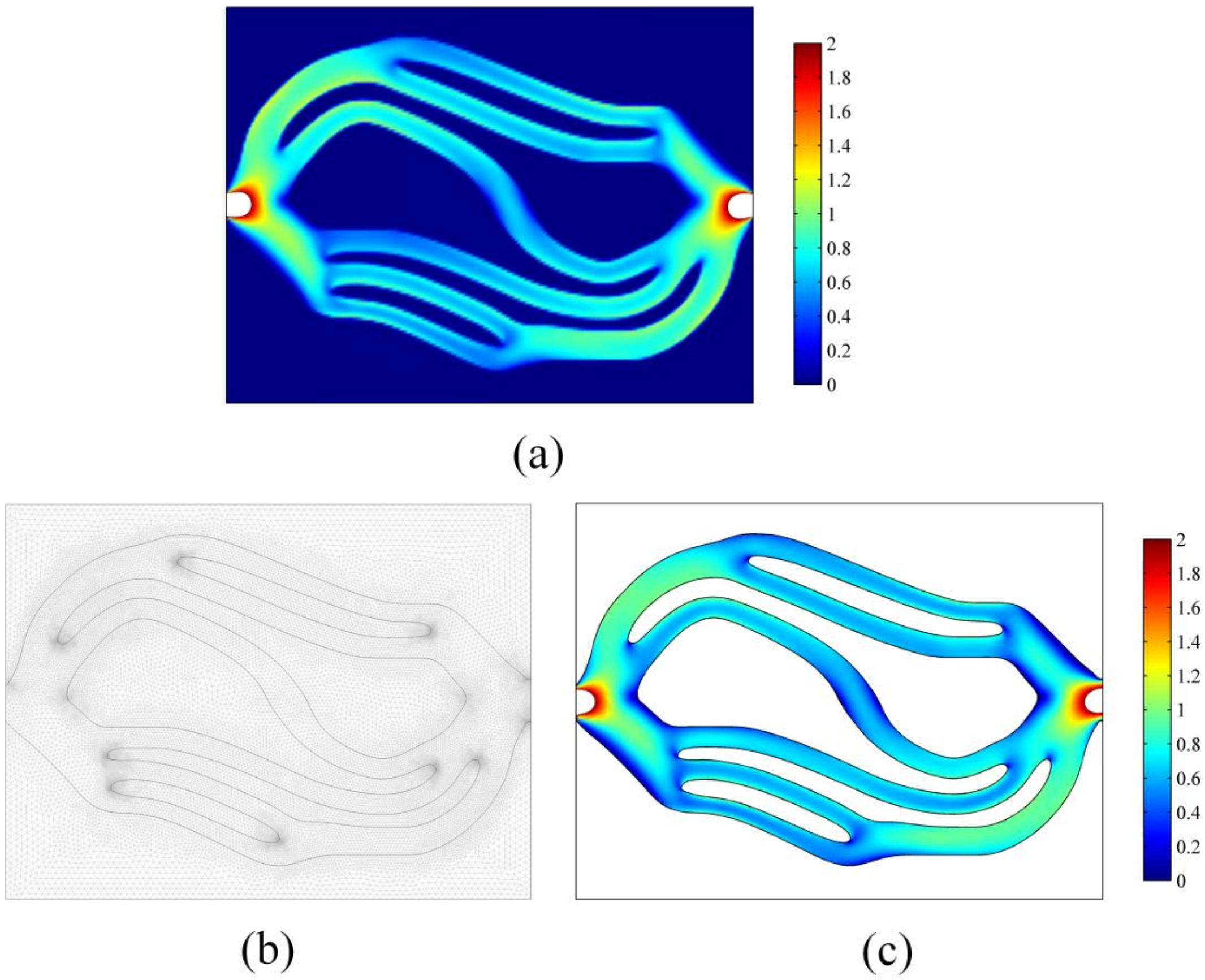

**Fig. 22.** Velocity-magnitude comparison using identical color limits: (a) calibrated diffuse-interface topology-optimization result; (b) body-fitted triangular mesh used for re-simulation; and (c) body-fitted result obtained by prescribing the corresponding tangential sliding-wall velocity.

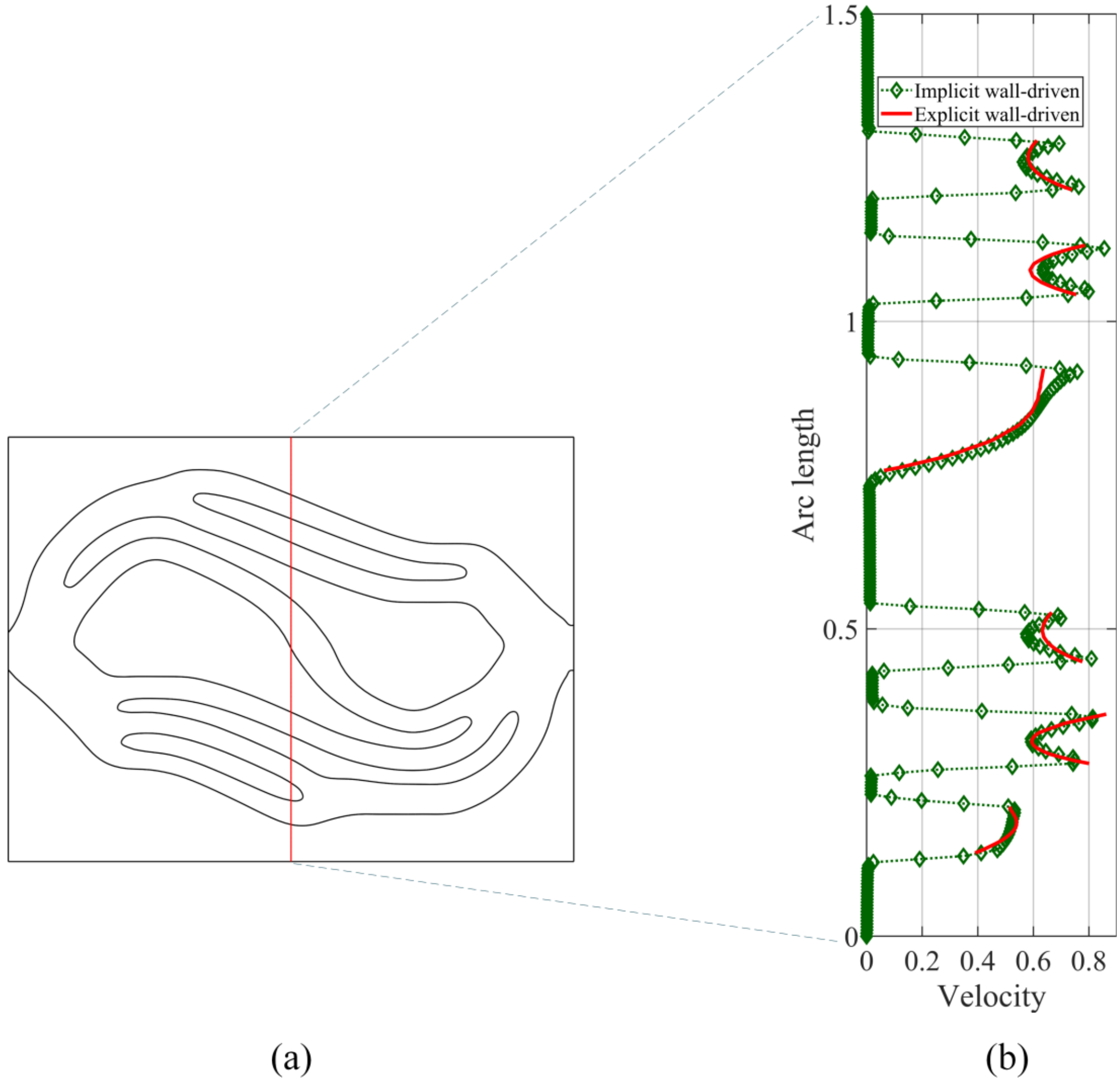

**Fig. 23.** Velocity-profile comparison for Fig. 22. (a) Sampling line at $x=1$; (b) diffuse-interface and body-fitted velocity profiles.

*4.4. Flux-constrained design of a 2D wall-driven flow splitter*

A two-dimensional splitter with one inlet and three outlets is used to examine flow-distribution control. The inlet and outlets in Fig. 24(a) are open, and the remaining external boundaries are no-slip. The prescribed vector in Fig. 24(b) uses $U_b=1$ and $\mathbf{d}=[1,0]^T$, driving the flow in the positive $x$ direction.

The objective maximizes the inlet flow rate subject to prescribed outlet ratios [15,17,79]. For outlet $k$, the constraint is

$$\left(\frac{\int_{\Gamma_{ok}} \mathbf{u}\cdot\mathbf{n}_{ok}\,\mathrm{d}\Gamma}{Q_{\text{in}} r_{ok}}-1\right)^2 \le \varepsilon^2, \qquad k=1,\ldots,N_o, \tag{44}$$

where $Q_{\text{in}}$ is the inlet flow rate, $\mathbf{n}_{ok}$ is the outward normal on outlet $\Gamma_{ok}$, $r_{ok}$ is its prescribed fraction of the inlet flow, and $\varepsilon=0.05$ is the tolerance. The initial continuation value is $q=80$ and $\theta=1/3$. A $200\times200$ finite-element mesh is used, and support centers are placed at every other node, yielding a $101\times101$ lattice.

Figs. 25–27 correspond to $(r_{o1},r_{o2},r_{o3})=(1/3,1/3,1/3)$, $(1/4,1/4,1/2)$, and $(1/6,1/3,1/2)$, respectively. Each case compares $R_j=0.12$ and $R_j=0.18$. The smaller radius permits greater local variation, whereas the larger radius produces smoother passages.

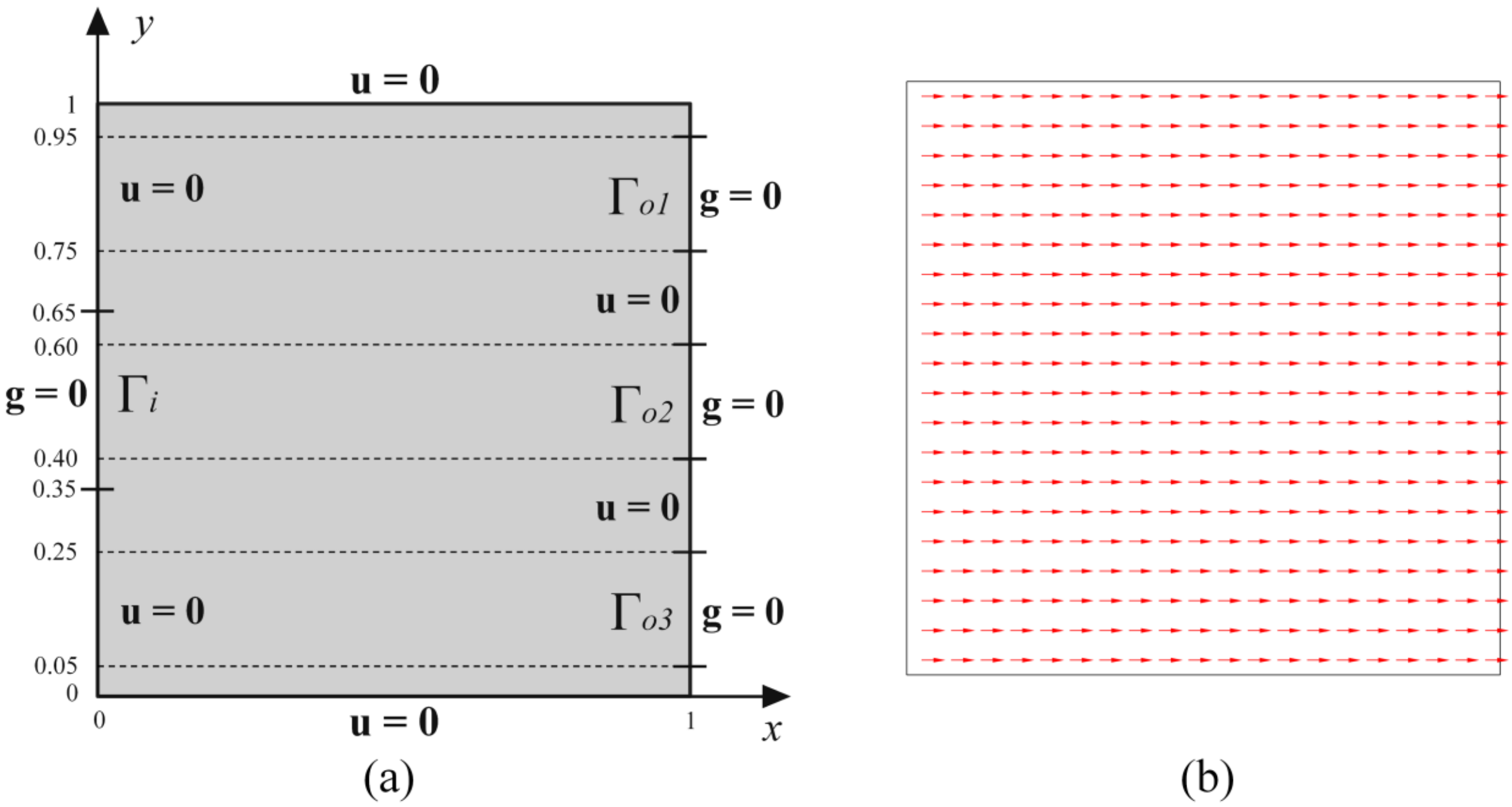

**Fig. 24.** Two-dimensional wall-driven splitter. (a) Design domain and boundary conditions; (b) prescribed $U_b\mathbf{d}$ with $U_b = 1$ and $\mathbf{d} = [1,0]^T$.

As shown in Figs. 25–27, the optimized networks adapt their branch capacities to the prescribed outlet flow rates. Comparable branches are formed for the equal-flow case, whereas increasing the demand at the bottom outlet produces a wider and more direct lower branch and a relatively thinner upper branch. The corresponding velocity fields confirm the intended redistribution of flow among the three outlets.

The phase-weighting factors in Figs. 25–27 remain localized near the optimized interfaces. The complete measure $\delta^{\beta}_{\Gamma,\phi}$ in Eq. (15) vanishes in the bulk fluid and solid phases.

A body-fitted re-simulation is performed for $R_j = 0.12$ and $(r_{o1}, r_{o2}, r_{o3}) = (1/6, 1/3, 1/2)$. Fig. 28 compares the two velocity fields. The main differences occur near junctions and narrow transition regions. The body-fitted outlet flow rates are 0.055, 0.112, and 0.163, satisfying the prescribed ratios.

Fig. 29 compares the velocity profiles along $x = 0.7$. The profiles agree closely in the main flow regions; differences are concentrated near channel boundaries and narrow gaps, where the diffuse-interface thickness and sharp-boundary reconstruction have the greatest influence.

These examples demonstrate outlet-ratio control in wall-driven splitters. The support radius $R_j$ controls geometric complexity, whereas Eq. (44) controls the outlet distribution. Sampling every other finite-element node reduces the number of CS-RBF design variables to about one quarter of the finite-element grid nodes without compromising interface resolution.

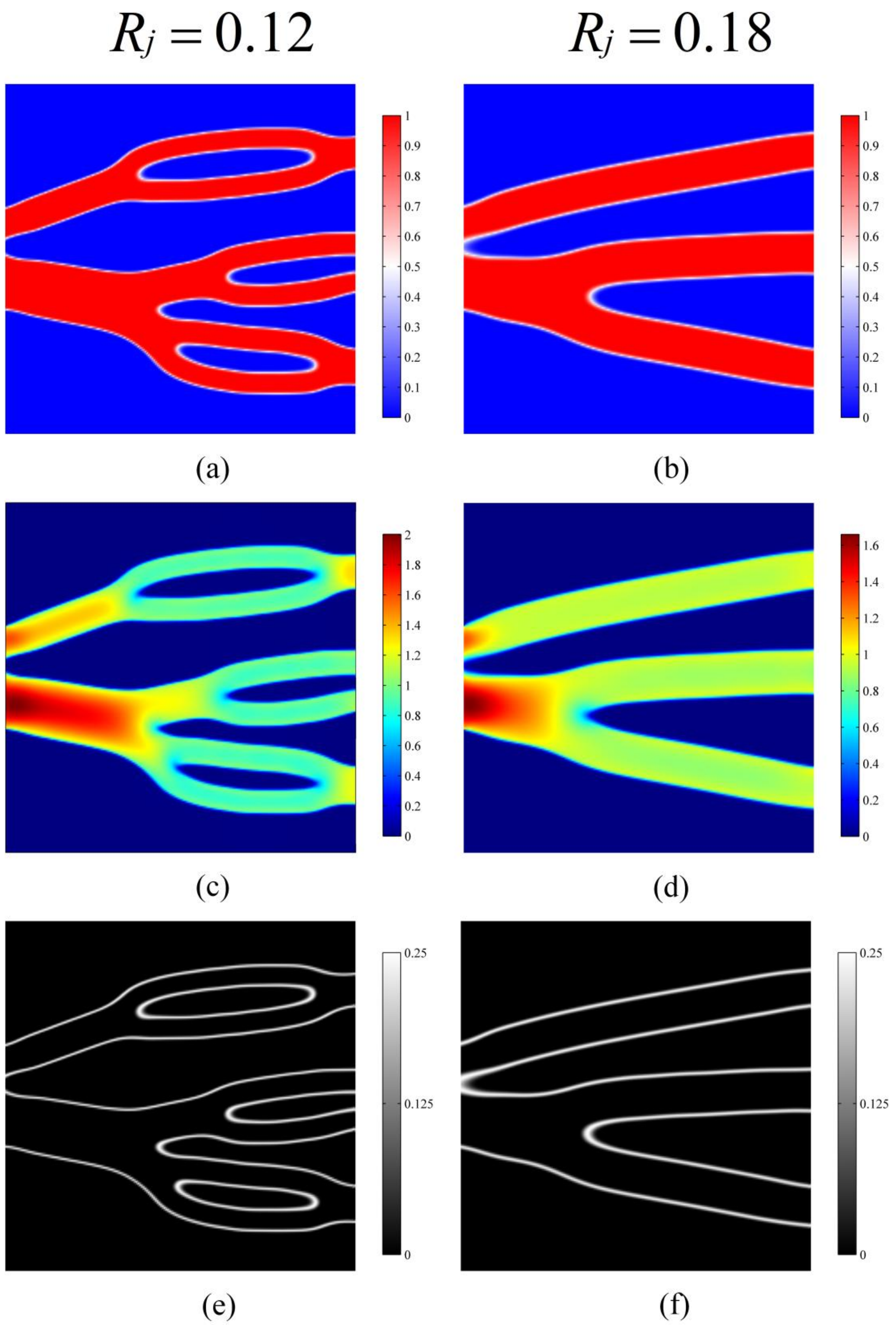

**Fig. 25.** Optimized results obtained with two CS-RBF support radii under the first prescribed outlet flow-rate distribution. Panels (a) and (b) show $\gamma$ ; panels (c) and (d) show the velocity magnitude; and panels (e) and (f) show the phase-weighting factor $\gamma(1-\gamma)$ .

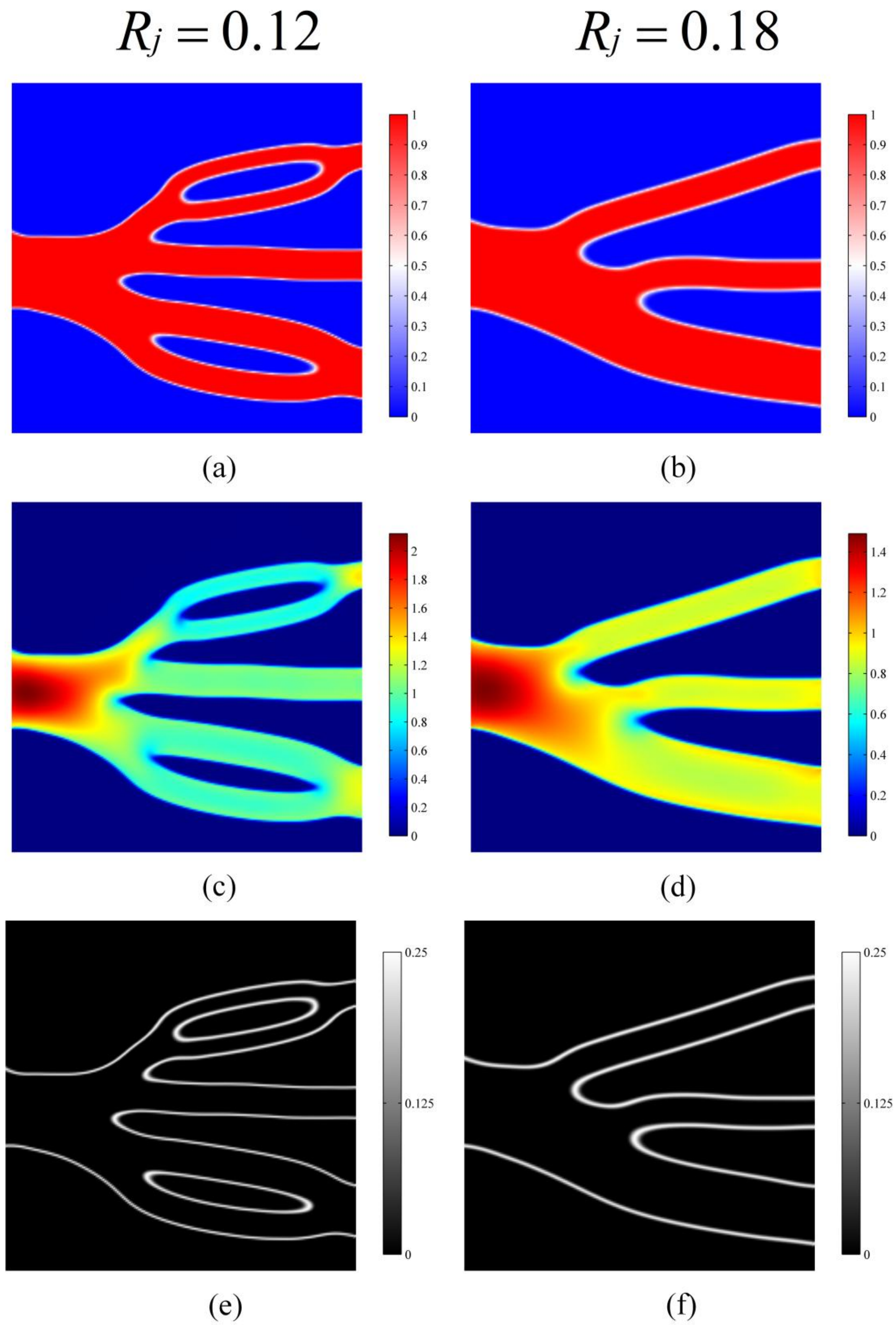

**Fig. 26.** Optimized results obtained with two CS-RBF support radii under the second prescribed outlet flow-rate distribution. Panels (a) and (b) show $\gamma$; panels (c) and (d) show the velocity magnitude; and panels (e) and (f) show the phase-weighting factor $\gamma(1-\gamma)$.

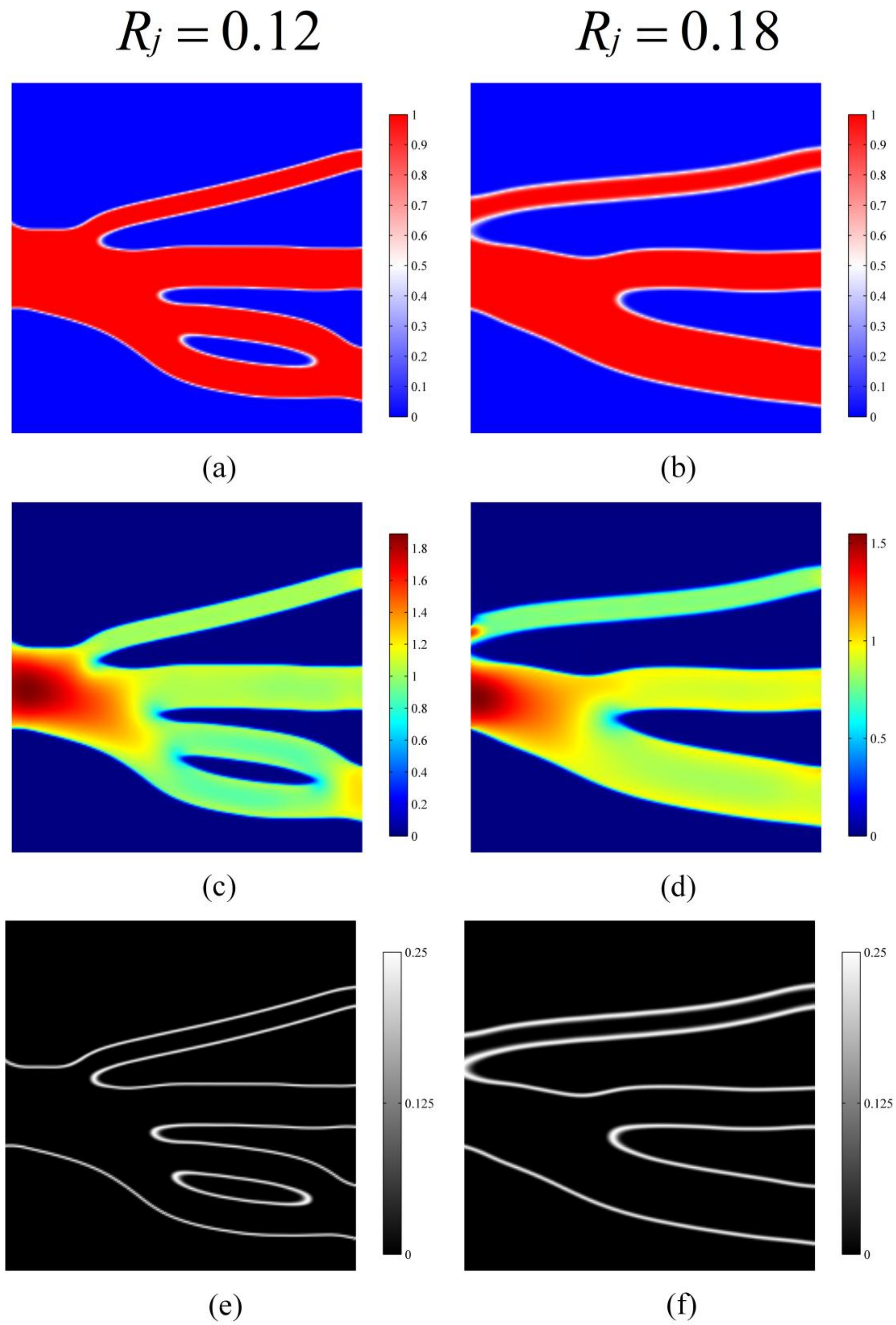

**Fig. 27.** Optimized results obtained with two CS-RBF support radii under the third prescribed outlet flow-rate distribution. Panels (a) and (b) show $\gamma$ ; panels (c) and (d) show the velocity magnitude; and panels (e) and (f) show the phase-weighting factor $\gamma(1-\gamma)$ .

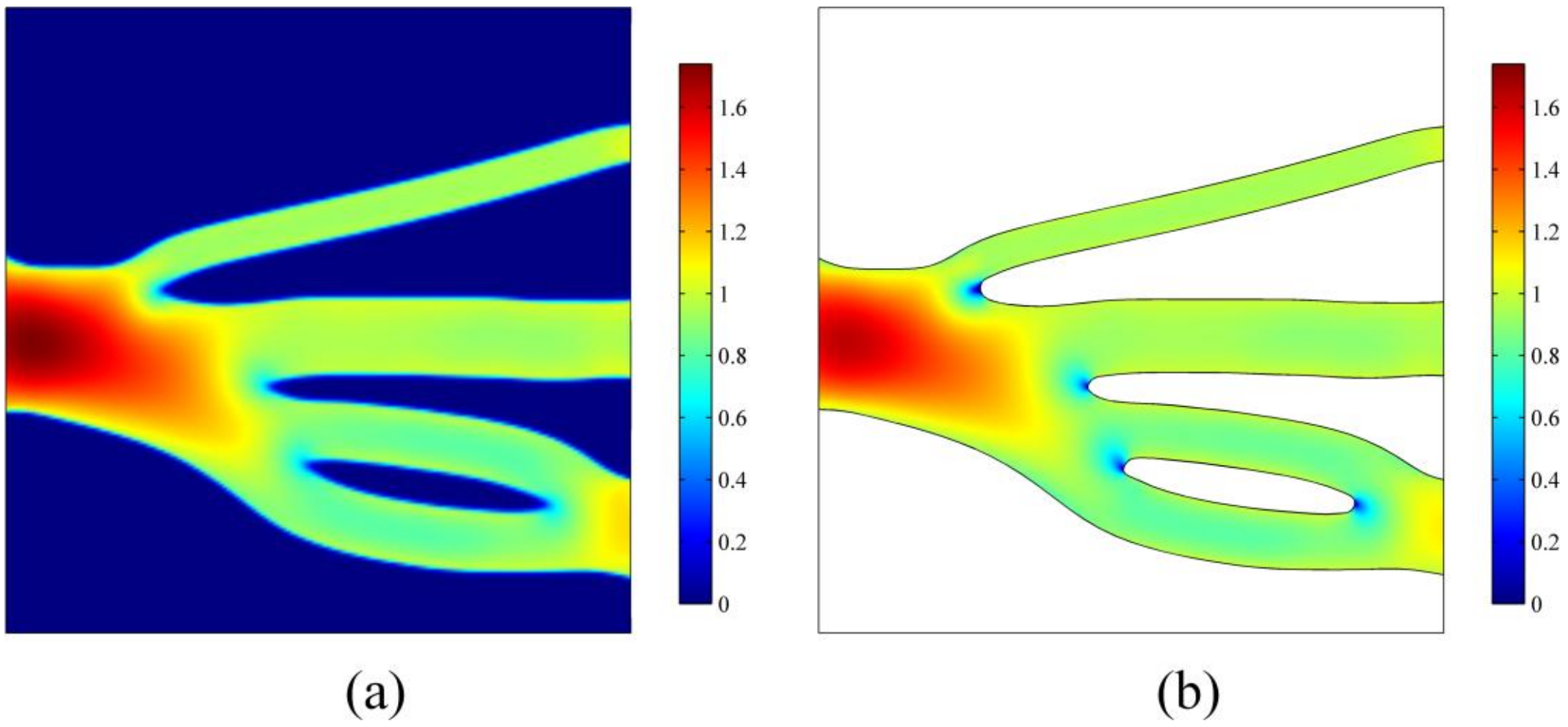

Fig. 28. Velocity-magnitude comparison for $R_j = 0.12$ and $(r_{o1}, r_{o2}, r_{o3}) = (1/6, 1/3, 1/2)$ : (a) diffuse-interface result; (b) body-fitted re-simulation.

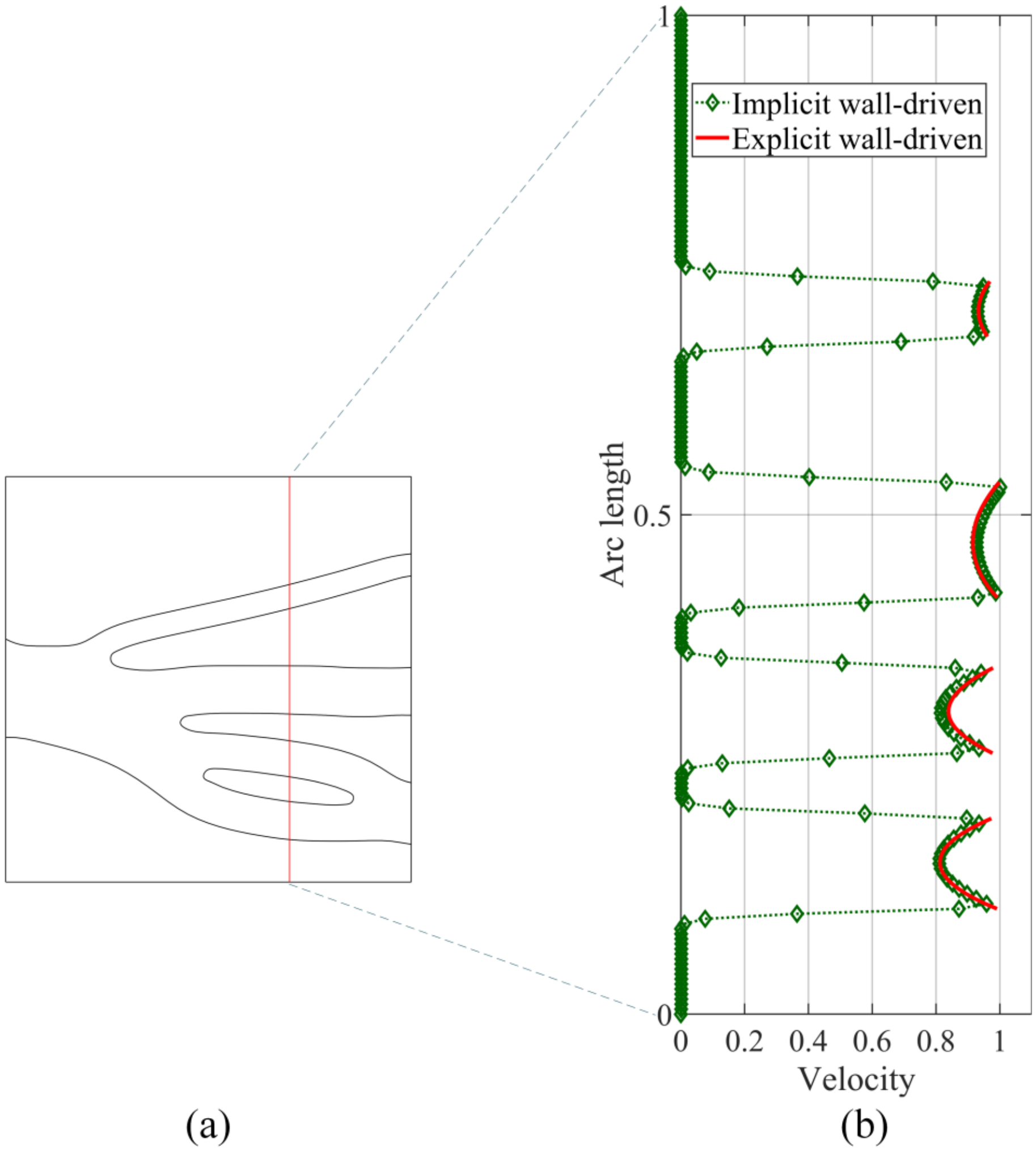

**Fig. 29.** Velocity profiles for Fig. 28. (a) Sampling line at $x = 0.7$ ; (b) diffuse-interface and body-fitted results.

*4.5. Flux-constrained design of a 3D wall-driven flow splitter*

The three-dimensional benchmark in Fig. 30 has one inlet and two outlets. The initial continuation value is $q = 80$ and the fluid-volume fraction is $\theta = 1/15$. The domain is discretized by 155,927 tetrahedral elements, and the CS-RBF support centers form a uniform $21 \times 11 \times 11$ lattice. The support radius is $R_j = 0.0625$, with $U_b = 1$ and $\mathbf{d} = [1, 0, 0]^T$.

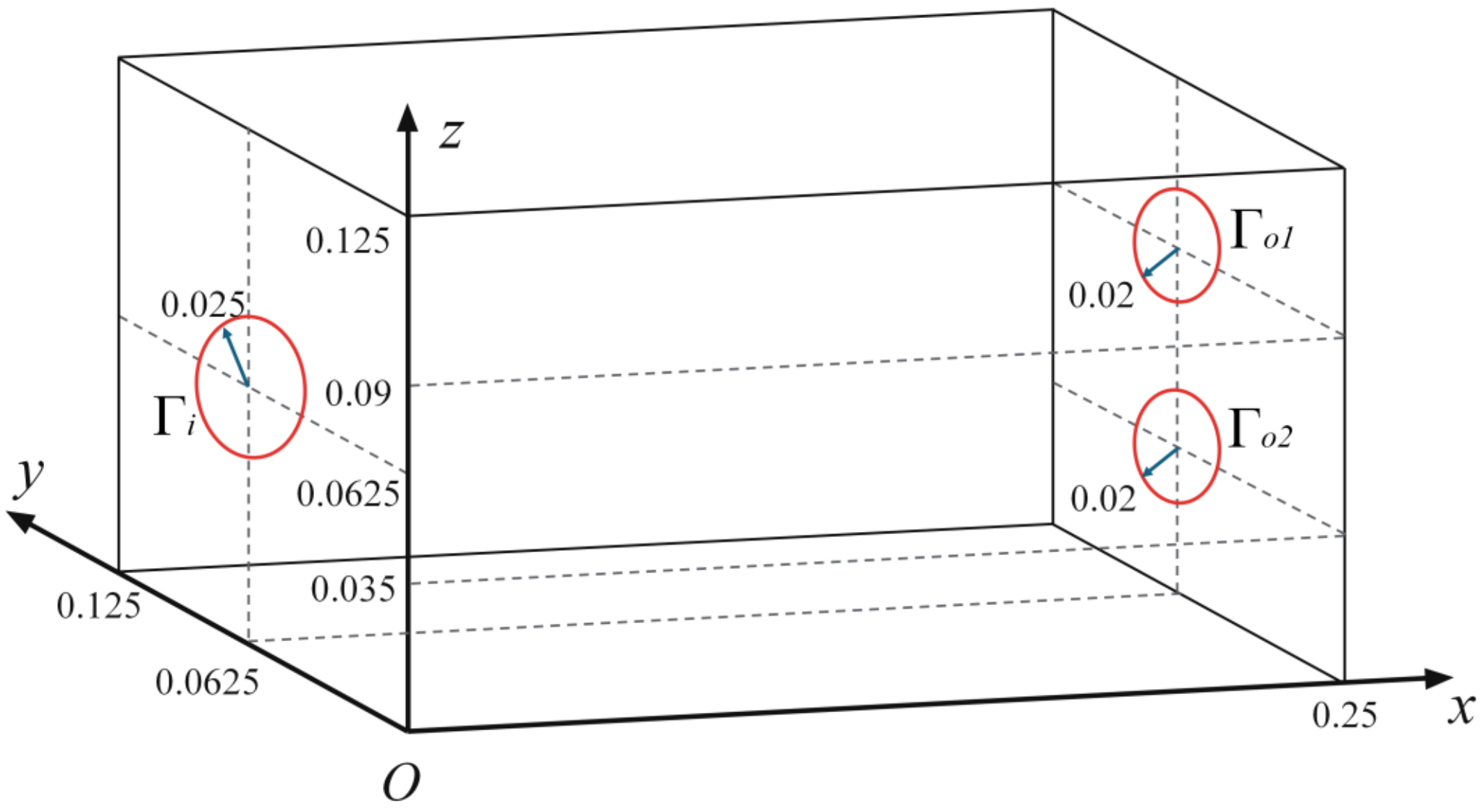

**Fig. 30.** Design domain for the 3D wall-driven flow splitter benchmark with one inlet and two outlets.

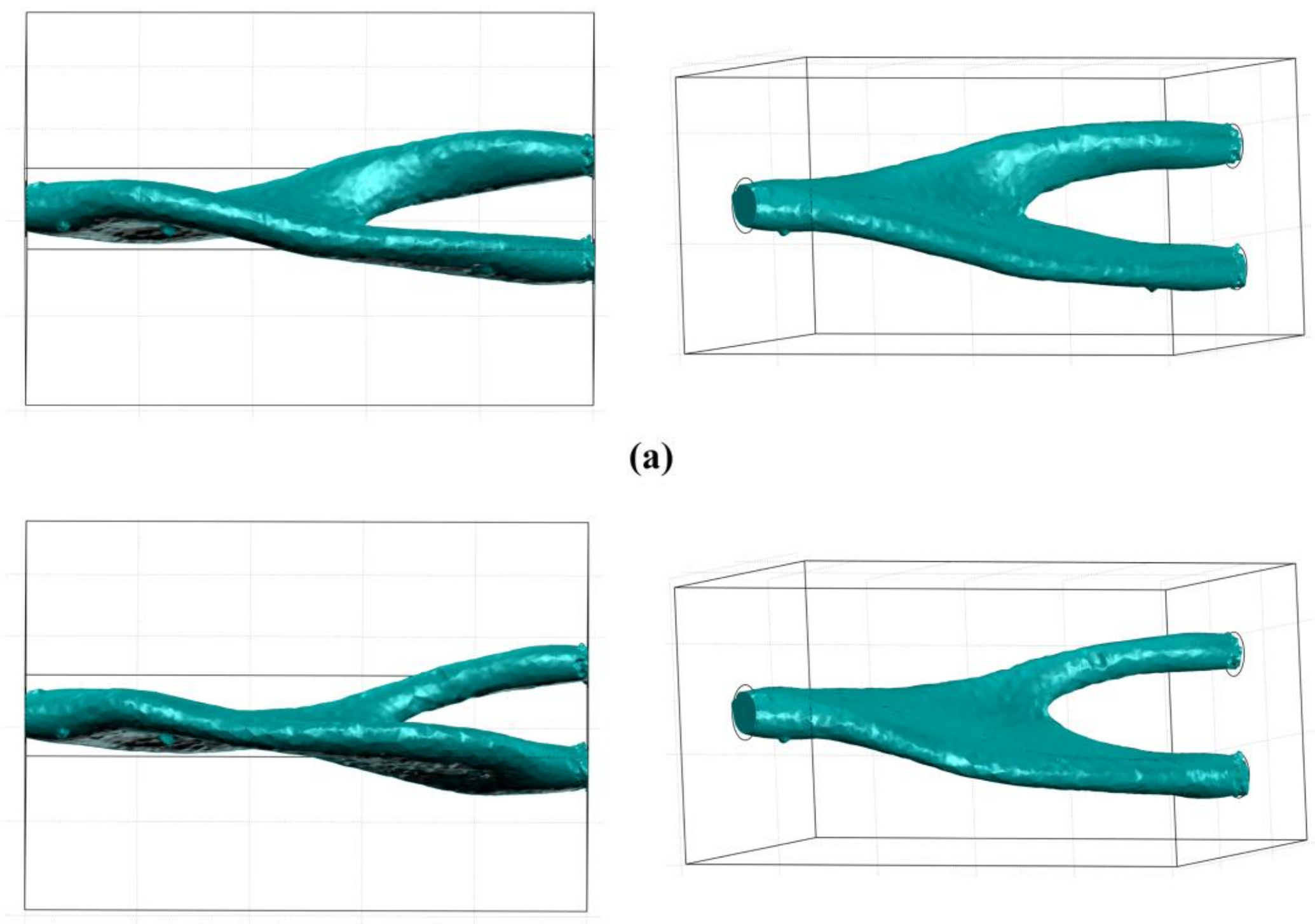

**(a)**

**(b)**

**Fig. 31.** Three-dimensional wall-driven splitters: (a) without a flux constraint; (b) with $(r_{o1}, r_{o2}) = (2/5, 3/5)$.

Fig. 31 compares the unconstrained design with the design satisfying $(r_{o1}, r_{o2}) = (2/5, 3/5)$. Both form continuous, smooth branching passages without isolated fragments.

A notable feature of the optimized three-dimensional channels is their flattened and twisted shape. The channels do not evolve into compact circular tubes. Instead, they become locally flattened and rotate gradually along the flow direction. This behavior is closely related to the wall-driven nature of the present problem. Under the same volume constraint, a flattened or twisted channel provides a larger wall surface area than a compact tube. Since the driving force is applied near the fluid–solid interface, the increased surface area enlarges the effective wall-actuation region and enhances the momentum transferred from the moving wall to the fluid. As a result, the optimizer tends to increase the surface-to-volume ratio of the channel to improve the total flow rate.

Without the flux constraint, the branch flows are determined by hydraulic resistance and available wall actuation. Imposing $(r_{o1}, r_{o2}) = (2/5, 3/5)$ changes the branch geometry to achieve the prescribed distribution.

The three-dimensional examples produced smooth branching passages with and without the outlet-flux constraint. The constrained design adjusted its branch geometry to satisfy the prescribed distribution. Only 2541

CS-RBF variables were used for a finite-element mesh containing 27,924 grid nodes, corresponding to 9.1% of the grid-node count.

### *4.6. Topology optimization of electroosmotic flow under the thin-EDL assumption*

This example applies the formulation to electroosmotic flow under the thin-EDL assumption. The electrolyte properties are $\rho = 1.0\times10^{3}\,\mathrm{kgm^{-3}}$ and $\mu = 1.0\times10^{-3}\,\mathrm{Pas}$. The relative permittivity is $\varepsilon_r = 80.2$, the conductivities are $\sigma_f = 0.12\,\mathrm{Sm^{-1}}$ and $\sigma_s = 1.0\times10^{-12}\,\mathrm{Sm^{-1}}$, and the zeta potential is $\zeta_0 = -0.1\,\mathrm{V}$. Table 3 lists the remaining parameters.

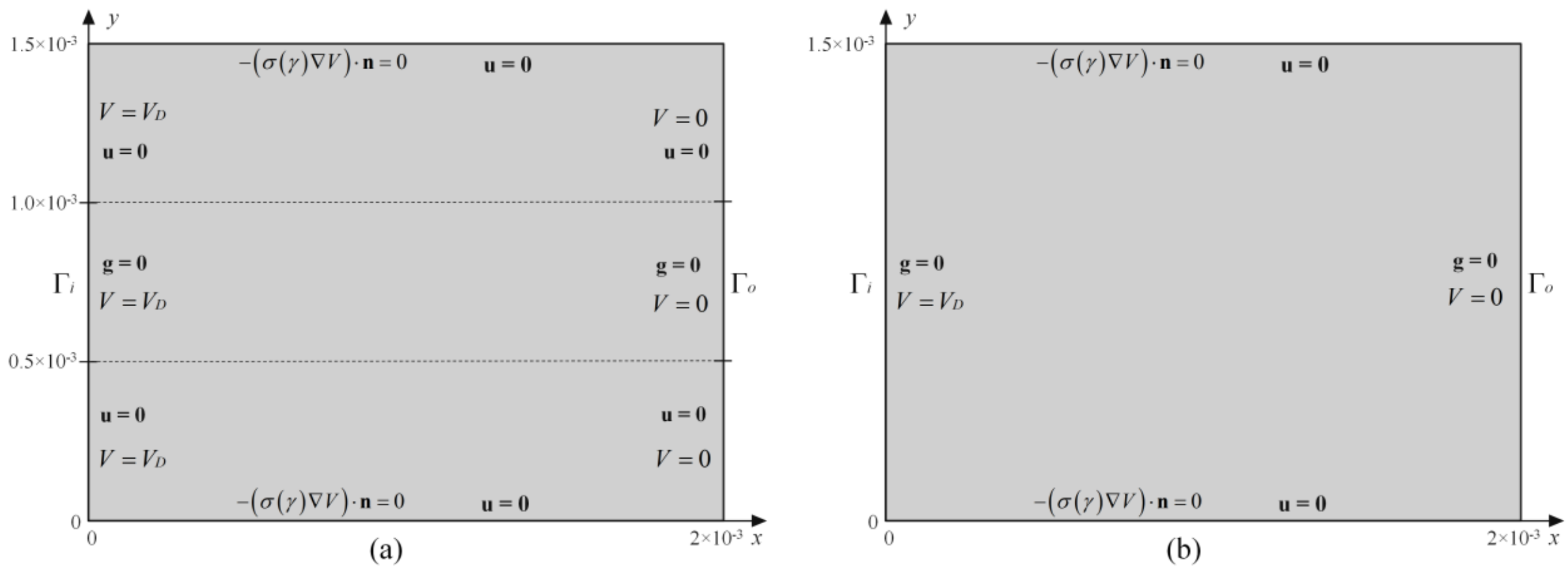

**Fig. 32.** Design domains and boundary conditions for the electroosmotic flow optimization problem. (a) Prescribed inlet and outlet segments located over the central one-third of the left and right boundaries, respectively; (b) fully open left and right boundaries without prescribed inlet or outlet segments.

For the first configuration, shown in Fig. 32(a), the inlet and outlet are restricted to the central one-third of the left and right boundaries, respectively. For the second configuration, shown in Fig. 32(b), the entire left and right boundaries are open. The inlet and outlet locations are therefore allowed to form freely as the topology evolves.

**Table 3**
Optimization parameters considered in the electroosmotic flow examples.

| $\alpha_{\max}$ | $\beta$ | $\kappa_T$ | Initial $q$ | $R_j$ | $\theta$ |
|---|---|---|---|---|---|
| $1\times10^{8}$ | 5 | $1.571\times10^{4}$ | 80 | $2\times10^{-4}$ | 1/3 |

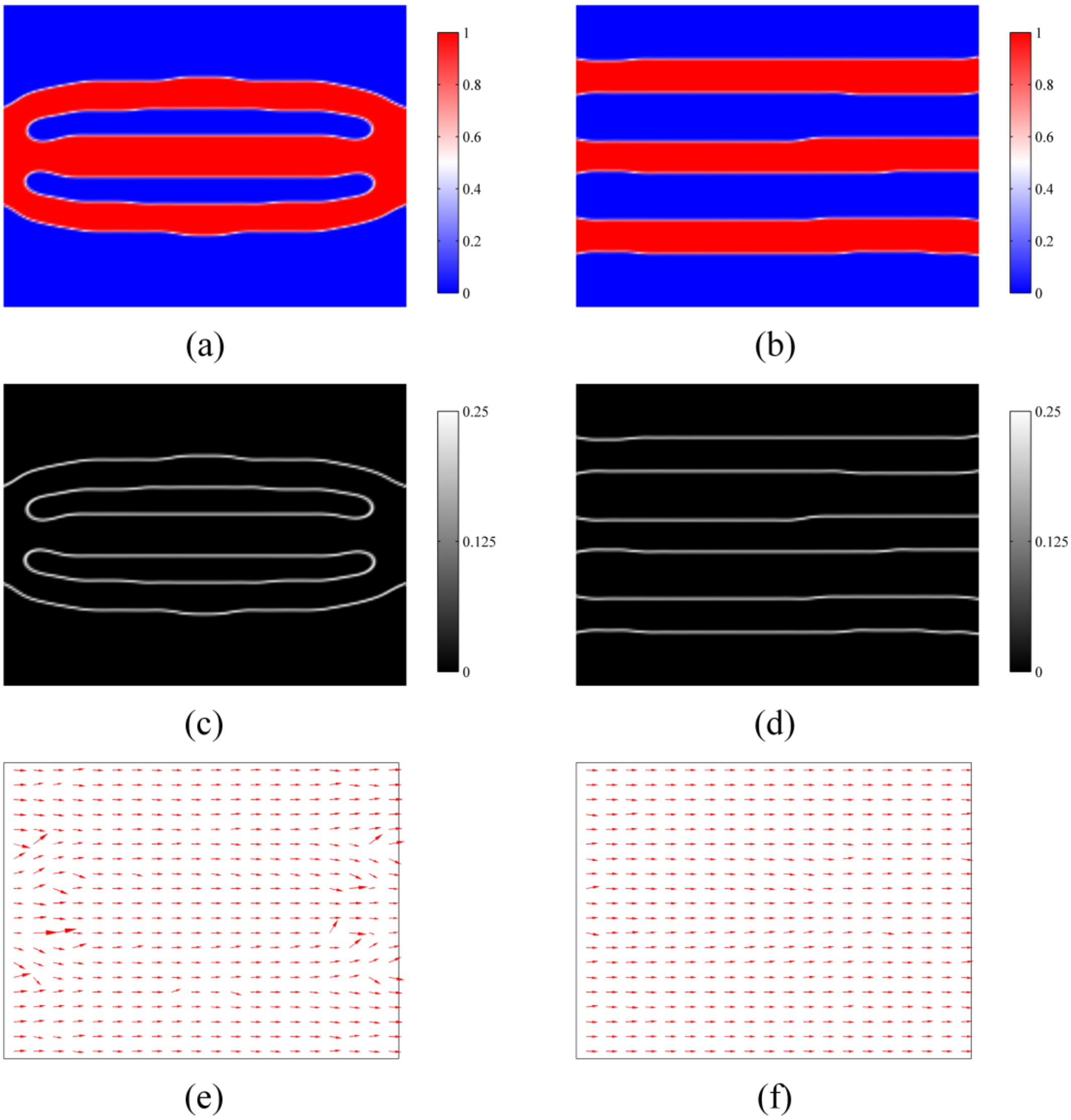

**Fig. 33.** Optimized electroosmotic designs. Panels (a) and (b) show $\gamma$ ; panels (c) and (d) show $\gamma(1-\gamma)$ ; panels (e) and (f) show $\mathbf{E}$ .

Fig. 33 shows the optimized electroosmotic designs for the two boundary configurations. With prescribed inlet and outlet segments, a multiply connected network forms around two elongated solid regions, increasing the fluid–solid interfacial length while maintaining connectivity. The electric field is predominantly streamwise and follows the curved passages near the ends of the solid regions. With fully open left and right boundaries, three nearly straight parallel channels are obtained, and the electric field remains approximately uniform and horizontal. In both cases, the channel walls tend to align with the local electric field, preserving a large tangential component for electroosmotic actuation. The multiple passages increase the effective actuation region under the fixed fluid-volume constraint, in contrast to pressure-driven designs, where additional wall area mainly increases viscous dissipation.

Figs. 33(c) and 33(d) show $\gamma(1-\gamma)$ . Together with $s_\phi$, it forms $\delta^{\beta}_{\Gamma,\phi}$ in Eq. (15), localizing electroosmotic momentum input to the transition layer.

The electroosmotic examples couple the evolving topology, design-dependent electric field, interface orientation, and Helmholtz–Smoluchowski slip velocity within the fixed-domain formulation. Both prescribed and freely formed inlet and outlet locations can be treated without changing the numerical framework.

## 5. Conclusions

This work presents a fixed-domain density-based topology optimization formulation for steady incompressible Navier–Stokes flows with design-dependent wall actuation. A continuous CS-RBF topological description function supplies the interface geometry, while its regularized projection defines the pseudo-density for Brinkman flow analysis. The prescribed wall velocity is represented by an equivalent volumetric momentum source localized through a normalized diffuse-interface measure. For a locally planar interface represented by uniformly distributed Wendland CS-RBFs, the transition width is related to the projection steepness and support radius, and the source coefficient is determined by balancing the momentum input with the Brinkman resistance at the interface center. Continuous adjoint analysis provides the design sensitivities.

The two- and three-dimensional examples show that wall-driven optimization produces topologies distinct from conventional inlet-driven designs. Under a fixed fluid-volume constraint, increasing the fluid–solid interfacial area enlarges the effective actuation region, resulting in separated or multiply connected two-dimensional networks and flattened, twisted three-dimensional passages. Prescribed outlet-flow distributions were achieved by corresponding changes in branch geometry. The CS-RBF support radius controls geometric complexity, and the parameterization reduced the number of design variables to approximately one quarter of the finite-element grid nodes in two dimensions and 9.1% in the three-dimensional example.

Body-fitted re-simulations reproduced the principal flow paths, velocity levels, and profile shapes predicted by the diffuse-interface model. The double-pipe benchmark gave a flow-rate difference of 3.59%, while the reconstructed splitter satisfied the prescribed outlet-flow distribution. The formulation was also applied to electroosmotic flow under the thin-EDL assumption. Prescribed inlet and outlet segments produced a multiply connected network produced a multiply connected network, whereas freely formed openings produced nearly straight parallel channels whose walls tended to align with the local electric field.

The present study is limited to steady incompressible flow and the thin-EDL electroosmotic model. Future work will consider other wall-actuation mechanisms and extensions to transient and turbulent flows.

## CRediT authorship contribution statement

**Limin Chen:** Writing – original draft, Software, Methodology, Investigation, Formal analysis, Data curation, Conceptualization. **Yuan Liang:** Writing – review & editing, Validation, Resources, Software, Investigation, Conceptualization. **Yongbo Deng:** Writing – review & editing, Visualization, Validation, Software, Data curation. **Chong Wang:** Writing – review & editing, Visualization, Validation, Data curation. **Youming Zhang:** Writing – review & editing, Supervision, Project administration, Funding acquisition, Conceptualization.

## Declaration of competing interest

The authors declare that they have no known competing financial interests or personal relationships that could have appeared to influence the work reported in this paper.

## Acknowledgements

This work was supported by the National Natural Science Foundation of China (Grant Nos. 52465005, 12572123, and 51875545).

## Data availability

Data will be made available on request.

## Appendix A. Characteristic transition width of the projected CS-RBF interface

This appendix derives the local width estimate reported in Eq. (16). The derivation assumes uniformly distributed two-dimensional Wendland supports with a common radius, an isolated locally planar interface, and support disks that are not truncated by the design-domain boundary.

The origin of the local coordinate system is placed at the center of the selected interface. The normal coordinate measures distance across the fluid–solid transition, whereas the tangential coordinate lies along the interface. Under the planar-interface and uniform-support assumptions, the convolved field is invariant in the tangential direction and can therefore be treated as a one-dimensional profile in the normal coordinate.

### A.1. Continuum representation

For the Wendland kernel adopted in this work, let $\chi = r / R_j$. A constant multiplicative factor is immaterial because the CS-RBF interpolation is normalized.

$$\psi(\chi)=\begin{cases}(1-\chi)^6(3+18\chi+35\chi^2), & 0\le\chi<1,\\ 0, & \chi\ge 1.\end{cases} \tag{A.1}$$

For a dense uniform support lattice, the normalized discrete interpolation is approximated by the continuum convolution below. The local coordinates $n$ and $t$ are normal and tangential to the planar interface, respectively, and $H(n)$ represents the corresponding half-space design.

$$\phi(\mathbf{x})=\frac{\int_{\mathbb{R}^2}\psi\left(\frac{\|\mathbf{x}-\xi\|}{R_j}\right)\lambda(\xi)\mathrm{d}\xi}{\int_{\mathbb{R}^2}\psi\left(\frac{\|\mathbf{x}-\xi\|}{R_j}\right)\mathrm{d}\xi},\qquad \lambda(n,t)=H(n). \tag{A.2}$$

In the continuum limit, the denominator is the kernel mass over one support disk, while the half-space distribution in the numerator separates supports assigned to the two phases. At the interface center, contributions from the two half-disks are equal. The remaining task is therefore to determine the slope of this normalized profile in the normal direction.

### A.2. Planar-interface gradient

The kernel moments required for the area normalization and the line intercepted through the interface center are

$$\int_0^1\chi\psi(\chi)\mathrm{d}\chi=\frac{1}{6},\qquad \int_0^1\psi(\chi)\mathrm{d}\chi=\frac{8}{9},\qquad C_R=2\pi R_j^2\int_0^1\chi\psi(\chi)\mathrm{d}\chi=\frac{\pi R_j^2}{3}. \tag{A.3}$$

The first kernel moment supplies the two-dimensional normalization through polar integration, whereas the second supplies the one-dimensional mass intercepted by the tangent line. Their ratio determines how rapidly the description field changes when the planar interface is displaced in its normal direction.

Differentiation in the normal direction converts the half-space step into a Dirac distribution on $n=0$. The remaining integral is therefore taken along the local tangent line, denoted by the coordinate $t$:

$$\left.\frac{\partial\phi}{\partial n}\right|_{n=0}=\frac{1}{C_R}\int_{-R_j}^{R_j}\psi\left(\frac{|t|}{R_j}\right)\mathrm{d}t=\frac{2R_j}{C_R}\int_0^1\psi(\chi)\mathrm{d}\chi. \tag{A.4}$$

Equation (A.4) follows by differentiating the half-space indicator. Its normal derivative is concentrated on the interface, leaving only the kernel integral along the tangent line. This is the central step that reduces the uniformly supported two-dimensional convolution to the local interface gradient.

For a locally planar interface, the description field varies only in the interface-normal direction n, while its tangential derivative vanishes. Hence, the gradient magnitude equals the magnitude of the normal derivative:

$$\|\nabla\phi\|=\left|\frac{\partial\phi}{\partial n}\right|. \tag{A.5}$$

Moreover, the gradient reaches its maximum at the interface center $n=0$. Substituting the two kernel moments into Eq. (A.4) therefore gives

$$s_\phi^*=\|\nabla\phi\|_\Gamma^*=\frac{16}{3\pi R_j}. \tag{A.6}$$

The maximum gradient is inversely proportional to the support radius. Larger supports smooth the topological description over a greater distance, whereas smaller supports permit a steeper and more localized transition. The result is independent of any constant scaling of the kernel because the interpolation is normalized.

### A.3. Conversion to a physical transition width

The compact regularized projection changes only over the description-field interval $\Delta\phi=1/\beta$. Linearization about the threshold, $\phi(n)\simeq\eta+s_\phi^* n$, locates the two ends of the projected layer as

$$n_\pm\simeq\pm\frac{1}{2\beta s_\phi^*}. \tag{A.7}$$

The distance between these locations is therefore

$$w_\Gamma^* = n_+ - n_- \simeq \frac{1}{\beta s_\phi^*} = \frac{3\pi R_j}{16\beta}. \tag{A.8}$$

The width in Eq. (A.7) is the distance required for the locally linear description field to traverse the complete active interval of the regularized projection. It is therefore the full projected transition width, rather than a half-width or a fitted contour thickness.

For a curved or nonuniform interface, the corresponding local estimate is

$$w_\Gamma(\mathbf{x}_\Gamma) \simeq \frac{1}{\beta \| \nabla\phi(\mathbf{x}_\Gamma) \|}. \tag{A.9}$$

Equation (A.7) gives the characteristic local width for an approximately planar interface represented by uniformly distributed, fully supported CS-RBFs. When these assumptions are not satisfied, the transition width can be evaluated directly from Eq. (A.8).

## Appendix B. Derivation of the equivalent wall-actuation coefficient

This appendix derives the coefficient used in Eqs. (17) and (18). The topological description function defines both the selected interface and its local tangent plane, while the diffuse source replaces the sharp wall momentum input on the fixed computational mesh.

### B.1. Sharp-interface actuation and Brinkman resistance

The fluid–solid interface is the selected isocontour of the topological description function. Its unit normal, tangential projector, and prescribed tangential wall velocity are

$$\Gamma = \{\mathbf{x} : \phi(\mathbf{x}) = \eta\}, \qquad \mathbf{n}_\phi = \frac{\nabla\phi}{\| \nabla\phi \|}, \qquad \mathbf{P}_\phi = \mathbf{I} - \mathbf{n}_\phi \mathbf{n}_\phi^T, \qquad \mathbf{u}_b = U_b \mathbf{P}_\phi \mathbf{d}. \tag{B.1}$$

The Brinkman resistance is interpolated as

$$\mathbf{f}_B = -\alpha(\gamma; q)\mathbf{u}, \qquad \alpha(\gamma; q) = \alpha_{\max} \frac{1-\gamma}{1+q\gamma}. \tag{B.2}$$

For a locally planar interface, let $s > 0$ point into the solid. Neglecting inertia and tangential pressure variation inside the thin penetration layer gives a one-dimensional Brinkman balance and its decaying solution:

$$-\mu \frac{\mathrm{d}^2 \mathbf{u}_t}{\mathrm{d}s^2} + \alpha_{\max} \mathbf{u}_t = \mathbf{0}, \qquad \mathbf{u}_t(s) = \mathbf{u}_\Gamma \exp\left(-\frac{s}{\ell_B}\right), \qquad \ell_B = \sqrt{\frac{\mu}{\alpha_{\max}}}. \tag{B.3}$$

Integration through the solid-side layer gives the resistance per unit interfacial area:

$$\int_0^\infty \alpha_{\max} \mathbf{u}_t(s)\mathrm{d}s = \sqrt{\mu\alpha_{\max}}\, \mathbf{u}_\Gamma. \tag{B.4}$$

The resistance in Eq. (B.4) acts against the interfacial tangential velocity. Applying an equal surface input associated with the prescribed wall velocity gives the net interfacial traction

$$\mathbf{t}_\Gamma = \kappa_\Gamma \left(\mathbf{u}_b - \mathbf{u}_\Gamma\right). \tag{B.5}$$

Since the interface-impedance coefficient defined in Eq. (B.4) increases without bound as the maximum inverse permeability grows, a finite limiting traction requires

$$\mathbf{u}_\Gamma \to \mathbf{u}_b. \tag{B.6}$$

Accordingly, the sharp-interface wall actuation can be written in distributional form as

$$\mathbf{f}_\Gamma = \kappa_\Gamma \delta_\Gamma \mathbf{u}_b, \tag{B.7}$$

where $\delta_\Gamma$ denotes the surface Dirac measure supported on $\Gamma$. This sharp-interface source is the reference quantity regularized by the diffuse-interface representation introduced next.

### B.2. Normalized diffuse representation of the surface source

For the smooth topological description function, the interface is the isocontour defined in Eq. (B.1). The coarea identity converts an integral over this contour into a domain integral:

$$\int_\Gamma g\,\mathrm{d}\Gamma = \int_\Omega g\,\delta(\phi-\eta)\|\nabla\phi\|\mathrm{d}\Omega. \tag{B.8}$$

Replacing the Dirac distribution by the compact phase weight $\gamma(1-\gamma)$ and imposing unit mass in the interface-normal direction gives

$$\hat{\delta}^{\beta}_{\Gamma,\phi} = \frac{\gamma(1-\gamma)}{I_\beta}s_\phi, \qquad I_\beta = \int_{\eta-1/(2\beta)}^{\eta+1/(2\beta)} \gamma(\phi)[1-\gamma(\phi)]\mathrm{d}\phi. \tag{B.9}$$

For the quintic projection used in this work, direct integration yields

$$I_\beta = \frac{25}{231\beta}, \qquad \hat{\delta}^{\beta}_{\Gamma,\phi} = \frac{231\beta}{25}\gamma(1-\gamma)s_\phi, \qquad \int_{-\infty}^{\infty}\hat{\delta}^{\beta}_{\Gamma,\phi}\,\mathrm{d}n = 1. \tag{B.10}$$

Thus, the integrated diffuse source is independent of the spatial width of a complete monotonic transition layer.

**B.3. Local matching at the interface center**

The final coefficient is obtained by matching the local actuation and Brinkman resistance at the center of the planar interface. At the final continuation value $q=1$, one has

$$\gamma = \frac{1}{2}, \qquad \alpha\left(\frac{1}{2};1\right) = \frac{\alpha_{\max}}{3}, \qquad \gamma(1-\gamma)s^*_\phi = \frac{1}{4}\frac{16}{3\pi R_j} = \frac{4}{3\pi R_j}. \tag{B.11}$$

Requiring the local principal tangential velocity to satisfy $\mathbf{u}=\mathbf{u}_b$ gives

$$-\alpha\left(\frac{1}{2};1\right)\mathbf{u} + \kappa_T\frac{4}{3\pi R_j}\mathbf{u}_b \simeq \mathbf{0}, \qquad \kappa_T = \frac{\pi\alpha_{\max}R_j}{4}. \tag{B.12}$$

Equivalently, the integrated Brinkman-layer estimate can be written with $\kappa_T^0 = (231\beta/25)\sqrt{\mu\alpha_{\max}}$ and a theoretically determined factor $C_\kappa = \kappa_T/\kappa_T^0$:

$$C_\kappa = \frac{25\pi R_j}{924\beta}\sqrt{\frac{\alpha_{\max}}{\mu}}, \qquad C_\kappa\kappa_T^0 = \frac{\pi\alpha_{\max}R_j}{4}. \tag{B.13}$$

Substitution into the momentum equation gives the final fixed-domain model:

$$\mathbf{f} = -\alpha(\gamma;q)\mathbf{u} + \frac{\pi\alpha_{\max}R_j}{4}\gamma(1-\gamma)s_\phi\mathbf{u}_b. \tag{B.14}$$

The matching relation is local and asymptotic; residual differences may remain between the diffuse-interface and body-fitted sharp-interface solutions.